\documentclass[showpacs,twocolumn,aps,preprintnumbers,letterpaper]{revtex4}
\usepackage{amsmath,amssymb}
\usepackage{epsfig}
\usepackage{graphicx}
\usepackage{amsmath}
\usepackage{slashed}
\usepackage{amsfonts}
\usepackage{epstopdf}

\newcommand{\be}{\begin{equation}}
\newcommand{\ee}{\end{equation}}
\newcommand{\bear}{\begin{eqnarray}}
\newcommand{\eear}{\end{eqnarray}}
\newcommand{\ba}{\begin{array}}
\newcommand{\ea}{\end{array}}

\def\be{\begin{eqnarray}}
\def\ee{\end{eqnarray}}
\def\bea{\be}
\def\eea{\ee}

\def\roughly#1{\mathrel{\raise.3ex\hbox{$#1$\kern-.75em%
\lower1ex\hbox{$\sim$}}}}

\begin{document}

\title{Heavy Holographic Exotics:\\
Tetraquarks as Efimov States}

\author{Yizhuang Liu}
\email{yizhuang.liu@sjtu.edu.cn}
\affiliation{Tsung-Dao Lee Institute, Shanghai Jiao University, Shanghai, 200240, China}

\author{Maciej A. Nowak}
\email{maciej.a.nowak@uj.edu.pl}
\affiliation{M. Smoluchowski Institute
of Physics and Mark Kac Center for Complex Systems Research,
Jagiellonian University, 30--348 Krak\'o{}w, Poland}

\author{Ismail Zahed}
\email{ismail.zahed@stonybrook.edu}
\affiliation{Department of Physics and Astronomy, Stony Brook University, Stony Brook, New York 11794-3800, USA}


\date{\today}
\begin{abstract}
We provide a holographic description of non-strange multiquark exotics as compact topological molecules
by binding heavy-light mesons to a tunneling configuration in D8-D$\bar 8$ that is homotopic to the vacuum state
with fixed Chern-Simons number. In the tunneling process,  the heavy-light mesons transmute
to  fermions. Their binding   is generic and arises from a trade-off  between the dipole attraction
induced by the Chern-Simons term and the U(1) fermionic repulsion. In the heavy quark limit, the open-flavor
tetraquark exotics  $QQ\bar q\bar q$ and $\bar Q\bar Q qq$, emerge as bound Efimov states in a degenerate multiplet
$IJ^\pi=(00^+ , 01^+)$  with  opposite intrinsic Chern-Simons numbers $\pm \frac 12$. The hidden-flavor tetraquark exotics such as $Q\bar Q q\bar q$, $QQ\bar Q\bar q$ 
and  $QQ\bar Q\bar Q$ as compact topological molecules are unbound. Other exotics are also discussed.
\end{abstract}



\maketitle

\setcounter{footnote}{0}


\section{Introduction}

Several experimental collaborations~\cite{BELLE,BESIII,DO,LHCb} have reported  new multiquark exotic states
such as the neutral $X(3872)$ and the charged $Z_c(3900)^\pm$  and $Z_b(10610)^\pm$, a priori
outside the canonical quark model classification. More recently, the LHCb~\cite{LHCb}
has reported new pentaquark states $P_c^+(4380)$ and $P_c^+(4450)$ through the
decays $\Lambda_b^0\rightarrow J\Psi pK^-, J\Psi p\pi^-$~\cite{LHCbx}, and five narrow and neutral excited $\Omega^0_c$
baryon states that decay primarily to  $\Xi_c^+K^-$~\cite{LHCbxx}.

Some of the reported hidden-flavor tetraquark exotics appear to be loosely bound hadronic molecules of two
heavy-light mesons~\cite{MOLECULES,THORSSON,KARLINER,OTHERS,OTHERSX,OTHERSZ,OTHERSXX,LIUMOLECULE},
although other explanations for their composition have been also suggested in~\cite{MANOHAR,RISKA,SUHONG}. 
The first estimates of the open-flavor and compact   tetraquark exotics were made in the context of the bag model~\cite{RICHARD}, and the
random instanton model with full chiral and heavy quark symmetry in~\cite{MACIEK2,MACIEK3}, in line with recent estimates using constituent quark models~\cite{KR}.

The reported pentaquark states with hidden charm
initially suggested in~\cite{MAREK}, have been addressed by many~\cite{MANY,PENTARHO,VENEZIANO,COBI} including
the newly reported neutrals $\Omega_c^0$ as discussed in~\cite{OMEGAC,EARLIER,SUMRULE,LATTICEC}.
Given the difficulty to track QCD in the infrared, it is not easy to identify  a first principle mechanism for the
formation of these multiquark states.

Most of the multiquark states reported so far  involve both heavy and light quarks but fall outside the realm
of the canonical quark model~\cite{KR}.  It is well established that the light quark sector of QCD exhibits spontaneous chiral symmetry
breaking, while the heavy quark sector is characterized by heavy quark symmetry~\cite{ISGUR}.
Both symmetries are at the origin of the chiral doubling suggested in heavy-light mesons~\cite{MACIEK,BARDEEN},
and confirmed experimentally in~\cite{BABAR,CLEOII}. It is therefore important that a theoretical approach
to the multiquark states should have manifest chiral and heavy quark symmetry, a clear organizational
principle in the infrared, and should address concisely the multi-body bound state problem.

The holographic principle in general~\cite{HOLOXX,HOLOXXX,HOLOXXXX},
and the D4-D8-D$\bar 8$ holographic set up in particular~\cite{SSX} provide a framework for addressing QCD in the
infrared in the double limit of a large number of colors and strong coupling $\lambda=g_{YM}^2N_c$. It is
confining and exhibits spontaneous  chiral symmetry breaking geometrically. In leading order
in $1/\lambda$,  the light meson sector is well described by an effective action on the fused D8-D$\bar 8$ branes
that is consistent with known effective theories~\cite{HIDDEN}.  The same setup can be minimally
modified to account for the description of heavy-light mesons as well, with full account of heavy
quark symmetry~\cite{LIUHEAVY}. Light and heavy-light baryons are dual to instantons and instanton-heavy meson bound states in
bulk~\cite{SSXB,SSXBB,CSLIGHT,KOJI,CSTHREE,LIUHEAVY,CHIN}, providing a concise approach to the multi-body bound state problem. 
In a way, the holographic construction provides a  geometrical realization of  the Skyrmion and its variants~\cite{SKYRME,PENTARHO,SKYRMEHEAVY},
without the shortcomings of the derivative expansion.
Alternative holographic models for the description of heavy hadrons have been
developed  in~\cite{FEWX,BRODSKY,COBIX} without the dual strictures of chiral and heavy quark symmetrty.


The organization of the paper is as follows: in section 2 we recall the geometrical
setup  for the derivation of the heavy-light  effective action  with two light and one
heavy flavor. We detail the heavy-light Dirac-Born-Infeld (hereafter DBI)  action and the particular  classical fields
of interest for the description of the holographic multiquark states. In section 3 we
derive explicitly  a class of O(4) tunneling configurations with fixed Chern-Simons (hereafter CS) 
number in D8-D$\bar 8$, that interpolate continuously between a 
unit topological charge (fermion) and zero topological charge (boson). We also derive their 
associated fermionic zero modes. In section 4, we detail how a heavy meson attached to the
tunneling configuration transmutes to a fermion. We also derive the pertinent Hamiltonian
on the moduli associated to the topological molecule formed of heavy mesons attached
to the O(4) tunneling configuration. In the heavy quark limit, the open-flavor and non-strange
tetraquarks  and hexaquarks 
are found to be bound Efimov-like states. The hidden-flavor tetraquarks are not bound.
In general, heavier exotics are not bound.
Our conclusions are in section 5. We provide two appendices for completeness.

\section{ Holographic heavy-light effective action}

\subsection{D-brane set up}

\begin{figure}[h!]
\begin{center}
 \includegraphics[width=6cm]{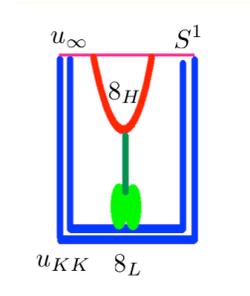}
  \caption{$N_f=2$ antipodal $8_L$ light branes, and one $8_H$  heavy brane shown in the $\tau U$ plane,
  with a bulk  O(4)  symmetric  tunneling configuration with a turning point,
  embedded in $8_L$ and a massive  HL-string connecting them.}
 \label{fig_branex}
 \end{center}
\end{figure}

The D4-D8-D$\bar 8$ setup for light flavor branes is standard~\cite{SSX}.  The minimal modification that accomodates
heavy mesons makes use of an extra heavy brane as discussed in~\cite{LIUHEAVY}. It consists
of $N_f$ light D8-D$\bar 8$ branes  (L) and one heavy (H) probe brane in the cigar-shaped geometry that
spontaneously breaks chiral symmetry.  A schematic description  of the set up for $N_f=2$ is shown in Fig.~\ref{fig_branex}.
We assume that the L-brane world volume consists of $R^{4}\times S^1\times S^4$ with
$[0-9]$-dimensions.  The light 8-branes are embedded in the $[0-3+5-9]$-dimensions and set
at the antipodes of $S^1$ which lies in the 4-dimension.
The warped $[5-9]$-space is characterized by a finite size  $R$ and a horizon at $U_{KK}$.

\subsection{DBI action}


The effective action on the probe L-branes
consists of the non-Abelian DBI   and CS  action.
After integrating over the $S^4$, the leading contribution in $1/\lambda$ to the DBI action is

\bea
\label{1}
S_{\rm DBI}\approx -\kappa\int d^4x dz\,{\rm Tr}\left({\bf f}(z){\bf F}_{\mu\nu}{\bf F}^{\mu\nu}+{\bf g}(z){\bf F}_{\mu z}{\bf F}^{\nu z}\right)
\eea
The warping factors are

\be
{\bf f}(z)=\frac{R^3}{4U_z}\,,\qquad {\bf g}(z)=\frac{9}{8}\frac{U_z^3}{U_{KK}}
\ee
with $U_z^3=U_{KK}^3+U_{KK}z^2$, and $\kappa\equiv a\lambda N_c$ and
$a=1/(216\pi^3)$~\cite{SSX}. Our conventions are $(-1,1,1,1,1)$ with $A_{M}^{\dagger}=-A_M$ and
the labels $M,N$ running over $\mu,z$ only in this section. All units are given in terms of  $M_{KK}=1$,  which
is readily recovered by dimensional inspection.
The effective fields in the field strengths are~\cite{LIUHEAVY}

\bea
\label{2}
&&{\bf F}_{MN}=\nonumber \\
&&\left(\begin{array}{cc}
F_{MN}-\Phi_{[M}\Phi_{N]}^{\dagger}&\partial_{[M}\Phi_{N]}+A_{[M}\Phi_{N]}\\
-\partial_{[M}\Phi^{\dagger}_{N]}-\Phi^{\dagger}_{[M}A_{N]}&-\Phi^{\dagger}_{[M}\Phi_{N]}
\end{array}\right)
\eea
The  matrix valued 1-form gauge field is
\be
\label{7}
{\bf A}=\left(\begin{array}{cc}
A&\Phi\\
-\Phi^{\dagger}&0
\end{array}\right)
\ee
For $N_f=2$, the naive Chern-Simons 5-form is

\be
\label{CSNAIVE}
S_{CS}=\frac{iN_c}{24\pi^2}\int_{M_5}\,{\rm Tr}\left(AF^2-\frac{1}{2}A^3F+\frac{1}{10}A^5\right)
\ee
We note that for only $N_f>2$ it
fails to reproduce the correct transformation law under the combined gauge and chiral transformations~\cite{CSLIGHT}.
In particular, when addressing the $N_f=3$ baryon spectra, (\ref{CSNAIVE}) does not treproduce the important hypercharge
constraint~\cite{CSLIGHT}, but can be minimally modified to do that.

For $N_f$ coincidental branes, the $\Phi$ multiplet is massless, but for separated branes
as illustrated in~Fig.~\ref{fig_branex} they are massive with the additional contribution

\bea
\label{8X3}
\frac 12 m_H^2 {\rm Tr}\left(\Phi^\dagger_M \Phi_M\right)
\eea
The value of $m_H$ is related to the separation between the light and heavy branes,
which is about the length of  the HL string. Below, $m_H$ will be taken as
the heavy meson mass.

\subsection{Light fields}

In the coincidental brane limit,
light baryons are interchangebly described as a flavor instanton or a D4 brane wrapping the $S^4$.
The instanton size is small with $\rho\sim 1/\sqrt{\lambda}$ after balancing  the order $\lambda N_c$
bulk gravitational attraction  with the subleading and of order $\lambda^0N_c$ U(1) induced topological
repulsion~\cite{SSX}.

To describe tetraquark states which carry zero topological charge or baryon number, but are still tightly bound by the
underlying light gauge field in holography, we suggest to use a tunneling configuration on the sphaleron path that is homotopic
to the vacuum state.  The configuration will carry fixed Chern-Simons number.
We will seek it using the maximally symmetric O(4) gauge field

\be
\label{XS3}
A_{M}(y)=-\overline{\sigma}_{MN}\partial_NF(y)\qquad \left.F_{zm}(y)\right|_{|y|=R}=0
\ee
subject to the condition of zero  $^\prime$electric$^\prime$ field strength  at the turning point $R=\rho$.
From here on $M,N$ runs only over $1,2,3,z$ unless specified otherwise.
If $\rho\sim 1/\sqrt{\lambda}$
is the  typical size of these tunneling configuations, then it is natural to recast the DBI action using the rescaling

\be
\label{S3}
&&(x_0, x_{M})\rightarrow (x_0,x_{M}/\sqrt{\lambda}), \sqrt{\lambda}\rho\rightarrow \rho\nonumber\\
&&(A_{0},A_M)\rightarrow (A_0, \sqrt{\lambda}A_M)
\ee
The rescaled fields  satisfy the equations

\be
\label{HL1}
D_{M}F_{MN}=0\qquad \partial_M^2A_0=-\frac 1{32\pi^2 a}F_{aMN}\star {F}_{aMN}
\ee
with the use of the Hodge dual notation, subject to the turning point condition (\ref{XS3})  in leading order in $1/\lambda$.
The detailed solution to (\ref{HL1}) will be given below.
Unlike the instanton which is stable, these
tunneling configurations are unstable and tend to relax to the vacuum state.  They are the O(4) analogue of
an instanton-anti-instanton configuration running to its  demise through  the valley. Below, we will show that they
can stabilize quantum mechanically when heavy mesons bind to them.

\subsection{Heavy-light fields}

Let  $(\Phi_0, \Phi_M)$ be the pair of heavy quantum fields that bind to the tunneling configuration above.
If again  $\rho\sim 1/\sqrt{\lambda}$ is their typical size, then it is natural to recast the heavy-light part of
the DBI action using the additional rescaling

\be
\label{S3S}
(\Phi_0,\Phi_M)\rightarrow (\Phi_0, \sqrt{\lambda}\Phi_M)
\ee
The interactions between the light gauge fields $(A_0, A_M)$ and
the heavy fields $(\Phi_0,\Phi_M)$  to quadratic order split to several contributions~\cite{LIUHEAVY}

\be
\label{RS1}
{\cal L}=aN_c\lambda {\cal L}_{0}+aN_c{\cal L}_{1}+{\cal L}_{CS}
\ee
which are quoted in the Appendix for completeness. Here we only need the leading contributions stemming
from (\ref{RS1}) in the additional heavy mass limit $m_H\rightarrow \infty$.  For that, we split
$\Phi_{M}=\phi_{M}e^{-im_Hx_0}$ for particles ($m_H\rightarrow -m_H$ for anti-particles). The leading
order contribution takes the form

\be
\label{RX66X}
{\cal L}_0=-\frac 12 \left|f_{MN}-\star f_{MN}\right|^2+2\phi_M^\dagger (F_{MN}-\star F_{MN})\phi_N\nonumber\\
\ee
subject to  the constraint equation $D_M\phi_M=0$ with

\be
f_{MN}=\partial_{[M}\phi_{N]}+A_{[M}\phi_{N]}
\ee
while the subleading contributions in (\ref{RS1}) to order $\lambda^0m_H$  simplify to

\be
\label{RX5}
&&\frac{{\cal L}_{1}}{aN_c}\rightarrow 4m_H\phi^{\dagger}_{M}iD_0\phi_{M}\nonumber\\
&&{\cal L}_{CS}\rightarrow \frac{m_H N_c}{16\pi^2}\phi^{\dagger}_{M}\star F_{MN}\phi_{N}
\ee

For self-dual light gauge fields with $F_{MN}=\star F_{MN}$,
the last contribution in (\ref{RX66X}) vanishes, and the minimum is reached for
$f_{MN}=\star f_{MN}$. This observation when  combined with the transversality condition for $D_M\phi_M=0$,
amounts to a first order equation for the combination $\psi=\bar \sigma_{M}\phi_{M}$ with $\sigma_M =(i, \vec \sigma)$, i.e.

 \be
 \label{RX66}
 \sigma_{M}D_{M}\psi= D \psi =0
 \ee
 as noted in~\cite{LIUHEAVY}. In a self-dual gauge configuration, the heavy spin-1 meson transmutes
 to a massless  spin-$\frac 12$ spinor that is BPS bound in leading order.  For the tunneling configuration in (\ref{XS3}) the
 self-duality condition no longer holds. With this in mind, we now proceed
to determine first the explicit tunneling configuration $(A_0, A_M)$
by solving (\ref{HL1}),  and then its variational  zero mode.

\section{Tunneling field and its fermionic zero mode}

In this section we detail the construction of a family of O(4) symmetric tunneling configurations (\ref{XS3})
solution to (\ref{HL1}). They carry fractional Chern-Simons number at the turning point, and interpolate parametrically between
the instanton and the sphaleron configuration at the turning point when
continued to Minkowski space. Their O(3) symmetric relatives through
a conformal transfornation are discussed in the Appendix. We note that similar configurations were used in the context
of explosive sphalerons and their applications to finite energy collisions~\cite{LS,EXPLO1,EXPLO2}.
We also derive their corresponding fermionic zero modes, which will prove useful for the discussion of the heavy-meson
bound states through transmutation.

\subsection{O(4) tunneling configuration}

\begin{figure}[h!]
\begin{center}
\includegraphics[width=6cm]{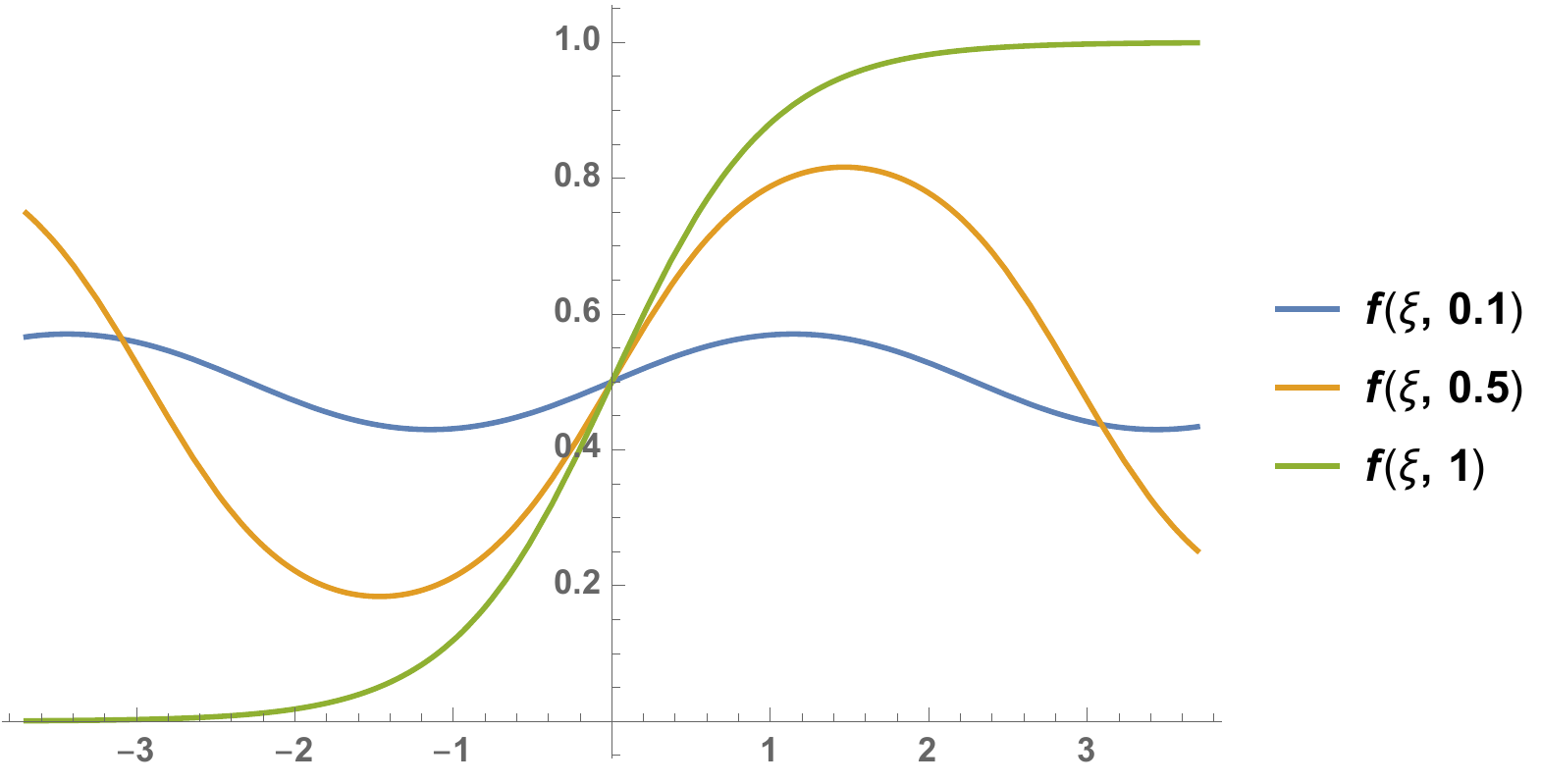}
  \caption{$f_k(\xi)$ for $k=0.1, 0.5, 1$.}
 \label{fig_fkx}
 \end{center}
\end{figure}

Consider the O(4)  static and symmetric
ansatz for the SU(2) gauge configuration

\be
A_{a M} (y) =&&2\,\bar\eta_{a MN}\,\frac{y_N}{y^2}f(\xi)
\label{Z1}
\ee
with the conformal variable $\xi(y)=\frac 12 {\rm ln} (y^2/\rho^2)$.
The anti-instanton and anti-sphaleron configurations follow from a similar
construction with a dual background $\bar\eta\rightarrow{\eta}$.
In terms of (\ref{Z1}) the static O(4) symmetric part of the Yang-Mills action in (\ref{1}) {\it without warping}
reads

\be
\label{Z2X}
S_{DBI}\rightarrow &&-{\kappa T}\int d^3xdz\,{F}_{aMN}{ F}_{aMN}\nonumber\\
=&&+\frac{24\pi^2 T}{g_\kappa^2}\int d\xi\left(\frac {f^{\prime 2}(\xi)}2 +V(f(\xi))\right)
\ee
with $T$ the length of time, the induced effective coupling
for the flavor gauge fields $g_\kappa^2=1/\kappa$, and  the double well potential

\be
\label{Z3X}
V(f)=2(f\bar f)^2
\ee
with $\bar f=1-f$.
The O(4) profile $f(\xi)$ extremizes (\ref{Z2X}) by satisfying

\be
\frac{d^2f}{d\xi^2} = 4(f^2-f) (2f-1)\,\,.
\label{Z3}
\ee
which is of the Jacobi type. Remarkably, the solution to (\ref{Z3}) with a sphaleron-like turning point  at $\xi=0$
with $f^\prime(\xi=0)=0$, can be found explicitly

\be
\label{Z4}
f_k(\xi)=\frac 12\bigg(1+\left(\frac{2k^2}{1+k^2}\right)^{\frac 12}{\rm sn}
\bigg(\xi \bigg(\frac 2{1+k^2}\bigg)^{\frac 12}, k\bigg)\bigg)\nonumber\\
\ee
with $\rm sn$ the Jacobi sine function. We note that the solution (\ref{Z4}) is $\xi$-{\rm periodic} with period

\be
\label{Z7}
T_k=2K(k)\left(\frac{1+k^2}2\right)^{\frac 12}
\ee
Here  $K(k)$ is the elliptic function,  and  $\xi \in [-\frac{T_k}{2},\frac{T_k}{2}]$.
In Fig.~\ref{fig_fkx} we show (\ref{Z4}) for $k=0.1, 0.5, 1$.

 The parameter $k$
 relates to the sphaleron-like energy at the turning point

\be
\label{Z5}
E_k=\frac{24\pi^2}{g_\kappa^2}V(f_k(\xi=0))=\frac{3\pi^2}{g_\kappa^2}\bigg(\frac{1-k^2}{1+k^2}\bigg)^2
\ee
At $k=0$ we recover the expected sphaleron energy  $E_0=3\pi^2/g_\kappa^2$ with the constant
profile $f_0(\xi)=\frac 12$. At the instanton point $k=1$ with zero energy $E_1=0$ we recover the
instanton interpolating profile

\be
\label{Z6}
f_1(\xi)=\frac{1}{2}+\frac{1}{2}{\rm sn}(\xi,1)=\frac{e^{2\xi}}{1+e^{2\xi}}
\ee
as illustrated in Fig.~\ref{fig_fkx}. In general, the solution
(\ref{Z4}) carries Chern-Simons number $N_{k}$ and  energy $E_k$
at the turning point, that are tied through the
profile of the potential (\ref{Z3X})

\be
\label{Z6X}
N_{k}(1-N_{k})= \frac 14 \left(\frac{E_k}{E_0}\right)^{\frac 12}
\ee
with $N_1=1$  the instanton topological charge, and $N_0=\frac 12$  the sphaleron Chern-Simons number.
Only the solution with $N_1=1$ is self-dual. All the other
configurations with $N_k<1$ are extrema rather than minima,  and therefore
prone to decay. They are homotopic to the vacuum state.

\begin{figure}[h!]
\begin{center}
 \includegraphics[width=6cm]{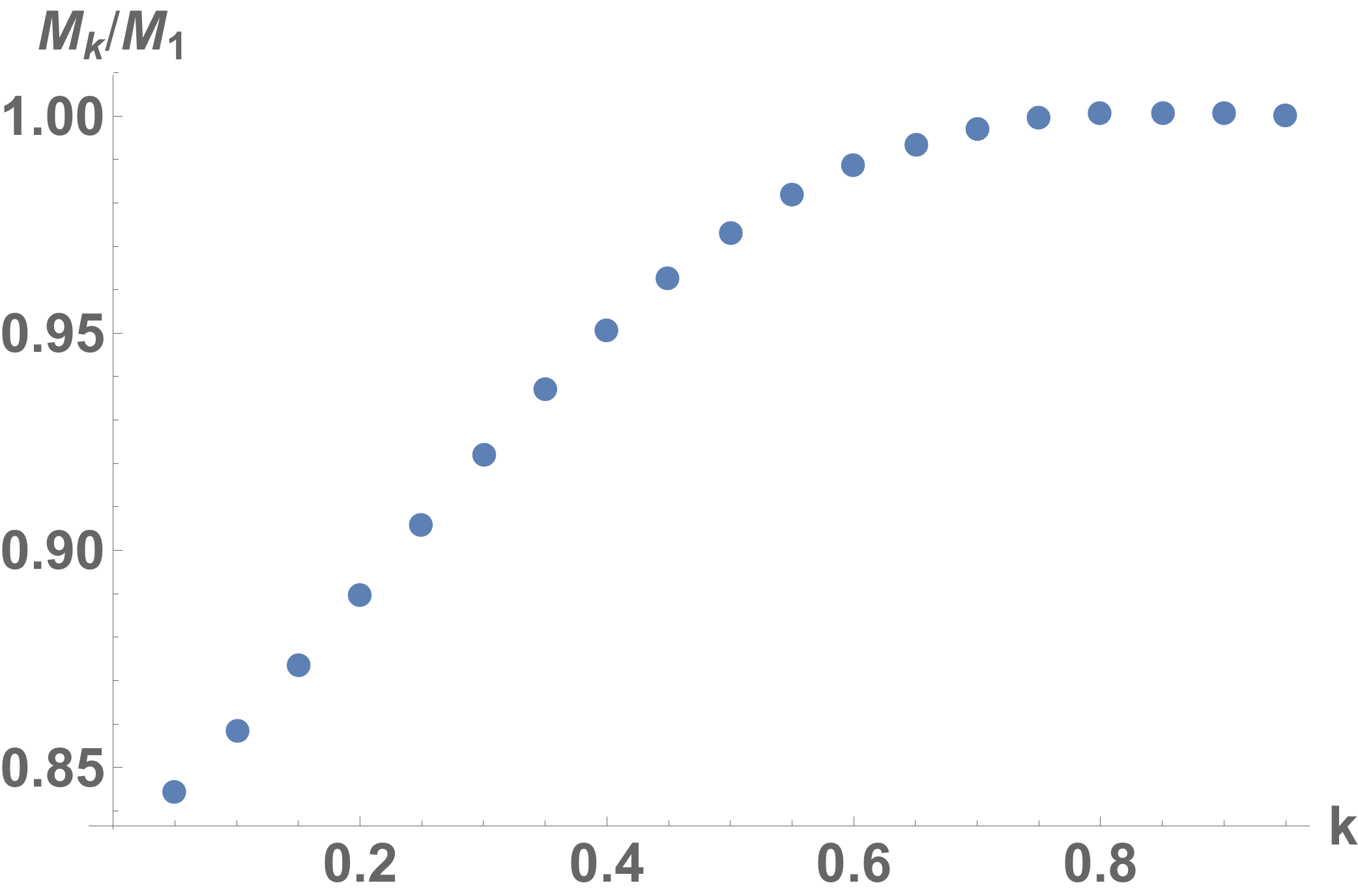}
  \caption{The mass ratio  $M_k/M_1$ versus k.}
 \label{fig_SkS1}
 \end{center}
\end{figure}

For the holographic dual  hadronic configurations, the more relevant quantity is the action
(\ref{Z2X}) for the generalized tunneling  configurations (\ref{Z4}). Since the solutions are periodic reflecting
on the periodicity of the sphaleron-ridge, we have for the fundamental period the action

\be
\label{Z8}
S_k=\frac{24\pi^2 T}{g_\kappa^2}\int_0^{T_k} d\xi\left(\frac {f^{\prime 2}(\xi)}2 +V(f(\xi))\right)
\ee
which gives  $S_1=(8\pi^2/g_\kappa^2)T$ at the instanton point as expected, and

\be
\label{Z9}
S_0=\frac{3\pi^2}{g_\kappa^2}T_0T=\frac {3\pi^3}{\sqrt{2} g_\kappa^2}T
\ee
at the sphaleron point. In particular, the {\it holographic} mass of  the tunneling configuration without warping, can be read from
(\ref{Z8}) as $M_k=S_k/{T}$. We note that the holographic mass ratio  at the sphaleron to instanton
point is $M_0/M_1=3\pi/8{\sqrt{2}}<1$.  In Fig.~\ref{fig_SkS1} we show  the mass ratio  $M_k/M_1=S_k/S_1$
for different values of $0\leq k\leq 1$.

\subsection{U(1) gauge field}

\begin{figure}[h!]
\begin{center}
 \includegraphics[width=6cm]{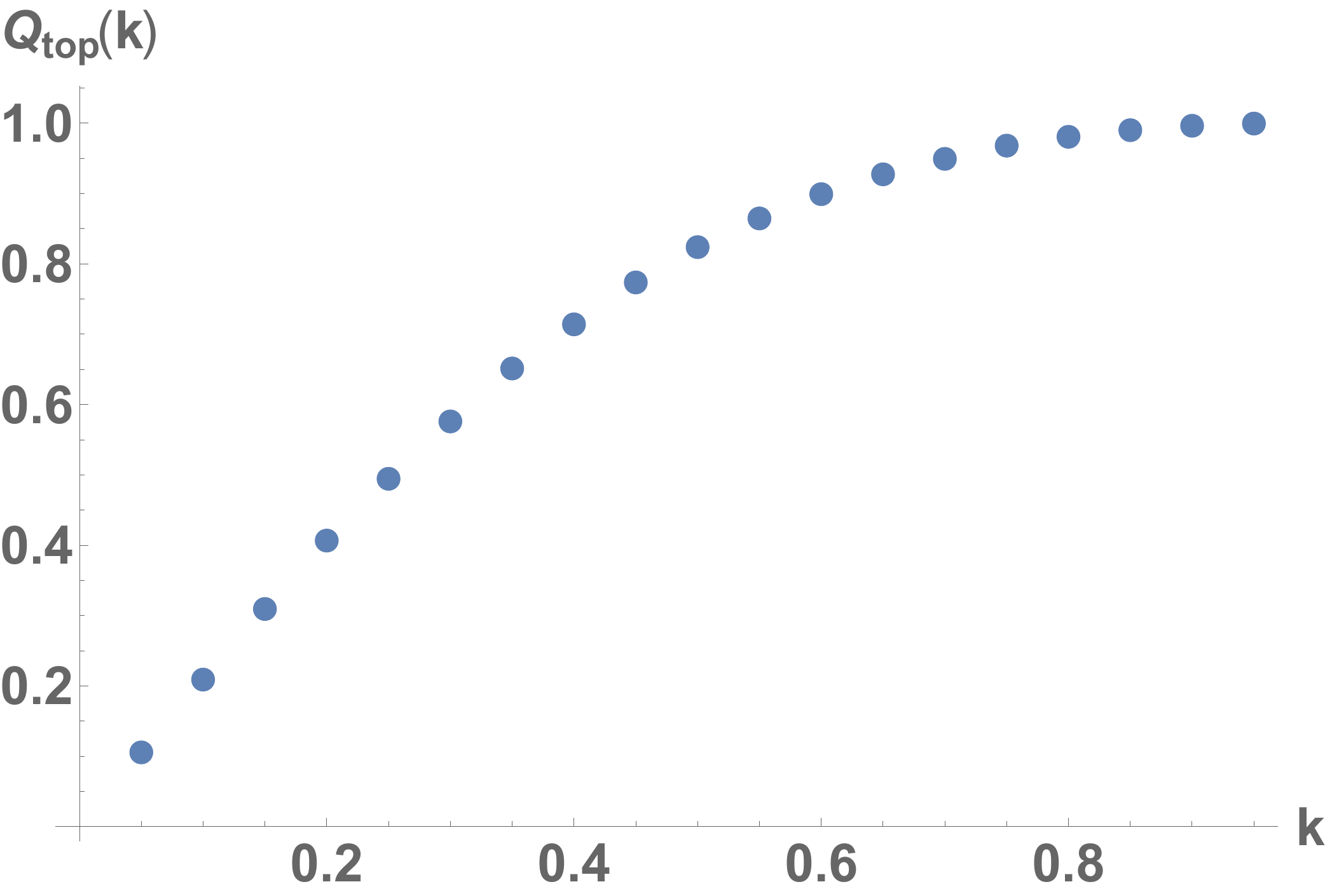}
  \caption{Topological charge  versus k.}
 \label{fig_Qtop}
 \end{center}
\end{figure}

The U(1) gauge  field in (\ref{HL1}) is sourced by the local topological charge density carried by the O(4)
tunneling configuration (\ref{Z1}). For that, it is useful to determine the field strength $F_{MN}$
and its dual $\star {F}_{MN}$, or more  explicitly the electric and magnetic fields

\begin{eqnarray}
\label{CL1}
y^2\vec{E}=&&\vec{\sigma}\left(f^{\prime}-\frac{(\vec{y})^2}{y^2}(f^{\prime}-2f\bar f)\right)\nonumber\nonumber \\
&&+\frac{(\vec{y}\cdot\vec{\sigma}\vec{y}-y_4\vec{\sigma} \times \vec{y})}{y^2}(f^{\prime}-2f\bar f)\nonumber\\
y^2\vec{B}=&&\vec{\sigma}\left(2f\bar f+\frac{(\vec{y})^2}{y^2}(f^{\prime}-2f\bar f)\right)\nonumber \\
&&-\frac{(\vec{y}\cdot\vec{\sigma}\vec{y}-y_4\vec{\sigma} \times \vec{y})}{y^2}(f^{\prime}-2f\bar f)
\end{eqnarray}
with $E^{i}=F^{i4}$ and $B^{i}=\frac 12{\epsilon^{ijk}}F^{jk}$. For self-dual fields
$f^\prime=2f\bar f$  and $\vec E=\vec B$ as expected for the instanton path. 
The U(1) field satisfies

\be
\label{CL2}
\partial_M^2A_0=-\frac{3}{4\pi^2 a y^4}\,\left(2f^\prime f\bar f \right)
\ee
which can be inverted if we define 

\be
A_0(y)=\frac 1{y^2}\,\phi_1(\xi)
\ee
so that $\phi_1(\xi)$  is solution to

\be
\label{PHI1}
\phi_1^{\prime \prime}-2\phi_1^{\prime}=&&
-\frac{3}{4\pi^2 a}\left(2f^\prime f\bar f \right)\nonumber\\
\equiv && F_0(\xi,k)\equiv -\frac 3{4\pi^2a}\,\mathbb{F}_0(\xi, k)
\ee
which is sourced by the topological charge density.
The net topological charge is

\be
\label{QTOP}
Q_{\rm top}(k)=3\int_{-\frac{T_k}2}^{\frac{T_k}2}d\xi \,\mathbb F_{0}(\xi, k)
\ee
Note that $\mathbb F_0(\xi, k)$  is a total derivative as it should,

\be
{\mathbb F}_0(\xi, k)=\left(f^2-\frac 23 f^3\right)^\prime
\ee
which is identically zero at the sphaleron point since  $f_0=\frac 12$. 
(\ref{QTOP}) 
is monotonous in $k$ as shown in Fig.~\ref{fig_Qtop}. It interpolates continuously between
the sphaleron path at $k=0$ with zero topological charge,  and the instanton path  at $k=1$ with
unit topological charge.

Eq.~(\ref{PHI1}) is readily solved using

\be
&&\phi_1(\xi)=C_1+C_2e^{+2\xi}\nonumber\\
&&+\frac{1}{2}\int_{-\frac{T_k}{2}}^{\xi}d\xi_1(e^{+2\xi-2\xi_1}-1)F_0(\xi_1,k)
\ee
 The constants of integration $C_{1,2}$ can be fixed by
choosing  the solution to satisfy the zero boundary condition at $\xi=-\frac{T_k}{2}$ and the regular boundary condition at $\xi=\frac{T_k}{2}$. This means that

\be
-C_2=e^{T_k}C_1=\frac{1}{2}\int_{-\frac{T_k}{2}}^{\frac{T_k}{2}}e^{-2\xi_1}F_0(\xi_1,k)d\xi_1
\ee
which explicitly gives

\be
\phi_1(\xi)=&&+\frac{1}{2}\int_{-\frac{T_k}{2}}^{\xi}d\xi_1(e^{+2\xi-2\xi_1}-1)F_0(\xi_1,k)\nonumber \\
&&-\frac{1}{2}(e^{2\xi}-e^{-T_k})\int_{-\frac{T_k}{2}}^{\frac{T_k}{2}}e^{-2\xi_1}F_0(\xi_1,k)d\xi_1\nonumber\\
\ee

\subsection{O(4) fermionic zero mode}

The zero mode solution in the O(4) configuration (\ref{1}) solves the Dirac equation

\be
\label{OZ1}
\left(\partial_M-iA_M\right)\gamma_M\,{\psi}=0
\ee
In the chiral basis with spin matrices $\bar\sigma_{Ms}=(1, -i\vec\sigma_s)$
 and

\begin{eqnarray}
\label{OZ2}
\gamma_5=
\left(
\begin{array}{cc}
  1 &   0  \\
 0  &   -1
\end{array}
\right)
\qquad
\gamma_\mu=
\left(
\begin{array}{cc}
  0 &   \sigma_{M s}   \\
  \bar\sigma_{Ms}  &   0
\end{array}
\right)
\end{eqnarray}
(\ref{OZ1}) splits into two chirality modes ${\psi}_{\pm}$ each conjugate of the other.
Note the  difference between the preceding conventions and the hermiticity of the
gauge field in this section only.
If we note that the t$^{\prime}$ Hooft symbol satisfies the color identity

\be
\label{OZ3}
\bar\sigma_{M c} \sigma_{N c}=\sigma_{ac} \overline{\eta}_{a MN}
\ee
with the color matrices  $\sigma_{Mc}=(1, -i\vec \sigma_c)$,
then the positive chirality mode associated to (\ref{OZ1}) satisfies

\be
\label{OZ4}
\left(\bar\sigma_{M s}\partial_M+\frac 12 \bar\sigma_{N s}\bar\sigma_{N c}\,\sigma_{M
c}\partial_M\,F\right)\,{\psi}_+ =0
\ee
with the spin and color matrices commuting, and $F(\xi)$  following from (\ref{Z1})

\be
\label{OZ4X}
F(\xi(y))=2\int_0^{\xi(y)}d\xi^\prime f(\xi^\prime)
\ee
Note that while writing  (\ref{OZ4}) we have added a U(1) part to the gauge field
for notational simplicity. It will be removed   in the final step below.
Eq.~(\ref{OZ4}) can be solved formally  using

\be
\label{OZ5X}
{\psi}_+=\varphi\,\chi_Q\qquad (\chi_Q)_{a\mu}=\epsilon_{a\mu}
\ee
wich is a singlet in color-spin space that satisfies

\be
\label{OZ5X1}
\sigma_{Ms}\chi_Q=\bar\sigma_{Mc}\chi_Q\qquad \bar\sigma_{Ms}\chi_Q=\sigma_{Mc}\chi_Q
\ee
Using (\ref{OZ5X}-\ref{OZ5X1}) in (\ref{OZ4}) yield

\be
\label{OZ5X2}
\left(\bar\sigma_{M s}\partial_M+\frac 12 \bar\sigma_{N s}\sigma_{N s}\,\bar\sigma_{M
s}\partial_M\,F\right)\,\varphi\chi_Q=0
\ee
It is here that we need to remove the U(1) contribution noted above through the substitution

\be
\label{OZ5X3}
\bar\sigma_{N s}\sigma_{N s}\,\chi_Q=
\left(1+(\vec\sigma_s)^2\right)\chi_Q\rightarrow (\vec\sigma_s)^2\chi_Q=3\chi_Q
\ee
leading to the O(4) symmetric equation for the zero mode amplitude

\be
\varphi^\prime+\frac 32 F^\prime \varphi=0
\ee
The spinor zero-mode of positive parity  is

\be
\label{OZ7}
{\psi}_+(y)= {\bf C}\,e^{-\frac 32 F(\xi (y))}\chi_Q\equiv f_0(y)\chi_Q
\ee
with the normalization constant fixed   within $T_k$

\be
\label{O28}
{\bf C}=\left|\int_{T_k} d^4y\, e^{-3F(\xi(y))}\right|^{-\frac 12}
\ee

\section{Multiquark exotics as topological  molecules}

In the triple limit of a large number of colors $N_c$, strong coupling $\lambda$ and heavy meson mass  $m_H$,
the holographic multiquark exotics  can be constructed by attaching to the O(4) tunneling gauge
configuration  an arbitrary number of heavy-light mesons.  Of course, in reality only a few can stick.
The fermionic repulsion induced through a U(1) coupling
to the Chern-Simons term stabilizes the tunneling configurations viewed as  an instanton-anti-instanton process.

\subsection{Heavy bound meson}

For $k=1$ with net topological charge 1,
the  heavy meson field in the self-dual classical background (\ref{XS3}) transmutes to a fermionic
zero mode (\ref{RX66}) as initially  noted in~\cite{LIUHEAVY}. For $k<1$ which is the case of interest
with fractional topological charge (fixed Chern-Simons number),  the classical background (\ref{XS3}) is no longer self-dual and the minimum
of (\ref{RX66X}) can be solved using a variational fermionic ansatz of the type (\ref{RX66})

\be
\label{O28X}
\phi_M\rightarrow \overline{\sigma}_M \psi_+\equiv \overline{\sigma}_M g(r)\chi_Q
\ee
with the radial coordinate $r=|y|$. In terms of (\ref{O28X}) the
leading contribution to the action ${\cal S}_0$ associated to  ${\cal L}_0$ in (\ref{RS1}) is

\be
{\cal S}_0=&&-\int d^4y\, \left(f^{\dagger}_{MN}f_{MN}+2\phi_M^{\dagger}F_{MN}\phi_N\right) \nonumber\\
=&&-\int 2\pi^2dr\,\left(6\mathbb S_0(g(r))\right)\,\chi_{Q}^{\dagger}\chi_Q
\ee
with

\be
\mathbb S_0(g)=r^3\left({g^\prime}^2+2g g^\prime G+g^2\left(-2G^\prime -\frac{6G}{r}+9G^2\right)\right)
\ee
Here $g^\prime ={dg}/{dr}$, $G(r)={f}/{r}$ and $G^\prime=dG/dr$ with the background gauge field expressed in the r-coordinate

\be
A_M=-\bar \sigma_{MN}\frac{y_M }{r}\,G(r)
\ee

The local minimum for $\mathbb S_0(g)$ requires that $g(r)$ satisfies

\be
g^{\prime\prime}+ \frac 35 g^\prime-\left(-3G^\prime -\frac{9G}{r}+9G^2\right)g=0
\label{EOM}
\ee
A special solution to (\ref{EOM}) is of the form

\be
\label{SPE}
g(r)=e^{-3\int_{0}^{r}dr^{\prime} G(r^{\prime})}=e^{-\frac{3}{2}F(\xi)}
\ee
which is readily seen to satisfy

\be
&&g^{\prime}+3Gg=0\nonumber\\
&&g^{\prime \prime}+3G^{\prime}g+3Gg^{\prime}=0
\ee
and therefore
\be
&&\Bigg(g^{\prime}+3Gg\Bigg)
+\Bigg(\frac{3}{r}-3G\Bigg)\Bigg(g^{\prime \prime}+3G^{\prime}g+3Gg^{\prime}\Bigg)
\nonumber \\
&& =g^{\prime \prime}+\frac{3}{r}g^{\prime}+\Bigg(-3G^{\prime}-\frac{9G}{r}+9G^2\Bigg)g=0
\ee
which is (\ref{EOM}).  We note that for $k=1$ which corresponds to the instanton path,
Eq.~(\ref{SPE}) reduces to the standard fermionic zero mode  ${(r^2+\rho^2)^{-\frac{3}{2}}}$.
(\ref{O28X}) with (\ref{SPE}) is transverse $D_M \phi_M=0$ for all values of $k$.
In~\cite{LIUHEAVY} $\chi_Q\rightarrow \chi_Q(t)$ describes the induced fermionic moduli upon binding,
which is how  the heavy quark of the original heavy-light meson manifests itself in this limit.

\subsection{Action  for the topological molecule}

With the above  in mind, and following
the arguments presented in~\cite{LIUHEAVY} (see section VB), the pertinent contributions to the
action  for the topological molecule to order $\lambda^0 m_H$ is

\be
\label{SZ3}
\frac{{S}_k}{aN_c}&&\approx\int dt \Bigg(\frac{M_k}{aN_c} -16m_H^2 \chi_Q^\dagger \chi_Q\Bigg)\nonumber\\
&&+\int dt\, d^4y\,\Bigg( 16m_H\,g^2(r)\,\chi_Q^{\dagger}i\partial_t\chi_Q\nonumber\\
&&+\frac{m_H}{8a\pi^2}\epsilon_{MNPQ}\phi^{\dagger}_{M}F_{NP}\phi_{Q}\Bigg)+S_C(A_0)
\ee
with the U(1) Coulomb contribution

\be
&&S_{C}(A_0=i\psi)=\nonumber\\
&&\int \left(\frac{1}{2}(\nabla \psi)^2+\psi\left(\rho_0[A]-16m_H g^2(r)\chi_Q^{\dagger}\chi_Q\right)\right)
\ee
due to the  attraction induced by the Chern-Simons  density $\rho_0(A)$ and the self-repulsion.
More explicitly, we have

\be
S_C(A_0)=&&16m_H\chi_Q^{\dagger}\chi_Q\int g^2(r)(-iA_0^{\rm cl})\nonumber\\
&&-\alpha_C(16m_H \chi_Q ^{\dagger}\chi_Q)^2
\ee
with the Coulomb factor

\be
\alpha_C=\frac{1}{2}\int g^2(r)\, \frac{1}{-\nabla^2} \,g^2(r)
\ee
The contribution to (\ref{SZ3})

\be
-2m_H(\phi_0 ^{\dagger}iD_M \phi_M-{\rm c.c})=0\nonumber
\ee
drops out  for any $k$,  thanks  to the transversality of the zero-mode (\ref{O28X}). We recall that
for $k=1$ the tunneling configuration is self-dual but not otherwise.
This configuration is an instanton with topological charge 1, and (\ref{SZ3})  describes  the
action of holographic heavy baryons~\cite{LIUHEAVY}.

\subsection{Hamiltonian for the topological molecule}

The molecular Hamiltonian associated to  (\ref{SZ3})  follows using the canonical rules for
$\chi_Q$ in the form

\be
\label{SZ4}
&&{\mathbb H}_{k}\approx  M_k +m_H\chi^\dagger_Q\chi_Q \nonumber\\
&&+\frac{\lambda\alpha_0(k)}{16 m_H\rho^2}\chi_Q^{\dagger}\chi_Q+\frac{3\alpha_1(k)}{4\pi^2a\rho^2}\chi_Q^{\dagger}\chi_Q +\frac{\alpha_2(k)}{8\pi^2aN_c\rho^2}(\chi_Q^{\dagger}\chi_Q)^2\nonumber\\
\ee
after the rescaling of the fermionic fields

\be
\chi_Q\rightarrow\frac 1{(16aN_cm_H)^{\frac 12}}\chi_Q
\ee
Switching to the conformal  coordinate $\xi$, and
using the explicit profile for the tunneling gauge configuration $f(\xi)$ in (\ref{Z4}), the U(1) gauge field $F_0(\xi, k)$
in (\ref{PHI1}) and the zero mode $g(\xi)$ in (\ref{SPE}),  we obtain for the k-dependent coefficients $\alpha_{0,1,2}(k)$

\be
\label{012}
\alpha_0(k)=&&\left({-12e^{+2\xi-3F}\left.f\right|_{\partial {\cal B}}}\right)\left({\int_{\cal B} d\xi e^{4\xi-3F}}\right)^{-1}\nonumber\\
\alpha_1(k)=&&\left({\int_{\cal B} d\xi e^{+2\xi-3F}\left(-\phi_1-\frac{1}{8}\left(\frac{f^{\prime}}{2}+\bar ff\right)\right)}\right)\nonumber\\
&&\times\left({\int_{\cal B} d\xi e^{4\xi-3F}}\right)^{-1}\nonumber\\
\alpha_2(k)=&&\left({\int_{\cal B} d\xi e^{+2\xi-3F}\phi_2}\right)\left({\int_{\cal B} d\xi e^{4\xi-3F}}\right)^{-2}\nonumber\\
\ee
As a reminder, the functions $\phi_{1,2}$ are explicitly given by

\be
\label{phi12}
&&\phi_1(\xi)=\frac{1}{2}\int_{-\frac{T_k}{2}}^{\xi}\,d\xi_1\,(e^{2(\xi-\xi_1)}-1)\,\mathbb F_0(\xi_1, k)\nonumber\\&&-\frac{1}{2}(e^{2\xi}-e^{-T_k})\int_{-\frac{T_k}{2}}^{\frac{T_k}{2}}\,d\xi_1e^{-2\xi_1}\mathbb F_0(\xi_1, k) \nonumber \\
&&\phi_2(\xi)= -\frac{1}{2}\int_{-\frac{T_k}{2}}^{\xi}\,d\xi_1\,(e^{2(\xi-\xi_1)}-1)\,e^{2\xi_1-3F(\xi_1, k)}\nonumber \\ &&+\frac{1}{2}(e^{2\xi}-e^{-T_k})\int_{-\frac{T_k}{2}}^{\frac{T_k}{2}}\,d\xi_1e^{-3F(\xi_1,k)}
\ee
We fix the  region of integration ${\cal B}$ to the period $T_k$ in (\ref{Z7})
or ${\cal B}=[-\frac{T_k}{2},\frac{T_k}{2}]$, with the boundary conditions
specified earlier. The integrals in (\ref{012})  for $\alpha_{1,2} (k)$ are not reducible to single
integrals upon the insertion of (\ref{phi12}) and integration by parts, since
$e^{2\xi-3F}$ does not integrate to a simple function.

The first contribution in (\ref{SZ4}) is the holographic mass (\ref{Z8}) of the tunneling configuration of order $\lambda N_c$, which is seen to reduce to the instanton mass (action) for $k=1$. The second contribution is the mass of the attached heavy quarks in  the mesonic molecule.
The contribution  linear in $\lambda$ stems from the boundary term for the quadratic heavy meson action. It is non-zero for $0\leq k<1$
due to the deviation of the tunneling gauge configuration from self-duality.
The $\alpha_1$ contribution stems from the U(1) Coulomb coupling of the charge $\chi^{\dagger}\chi$ to the background charge $\rho_0$ and the Chern-Simons terms.The $(\chi^{\dagger}\chi)^2$ contribution stems from the U(1)  Coulomb-like self-interaction. All of these contributions are similar to the $k=1$ case with the exception of the $\alpha_0$-contribution.

\begin{figure}[h!]
\begin{center}
 \includegraphics[width=8cm]{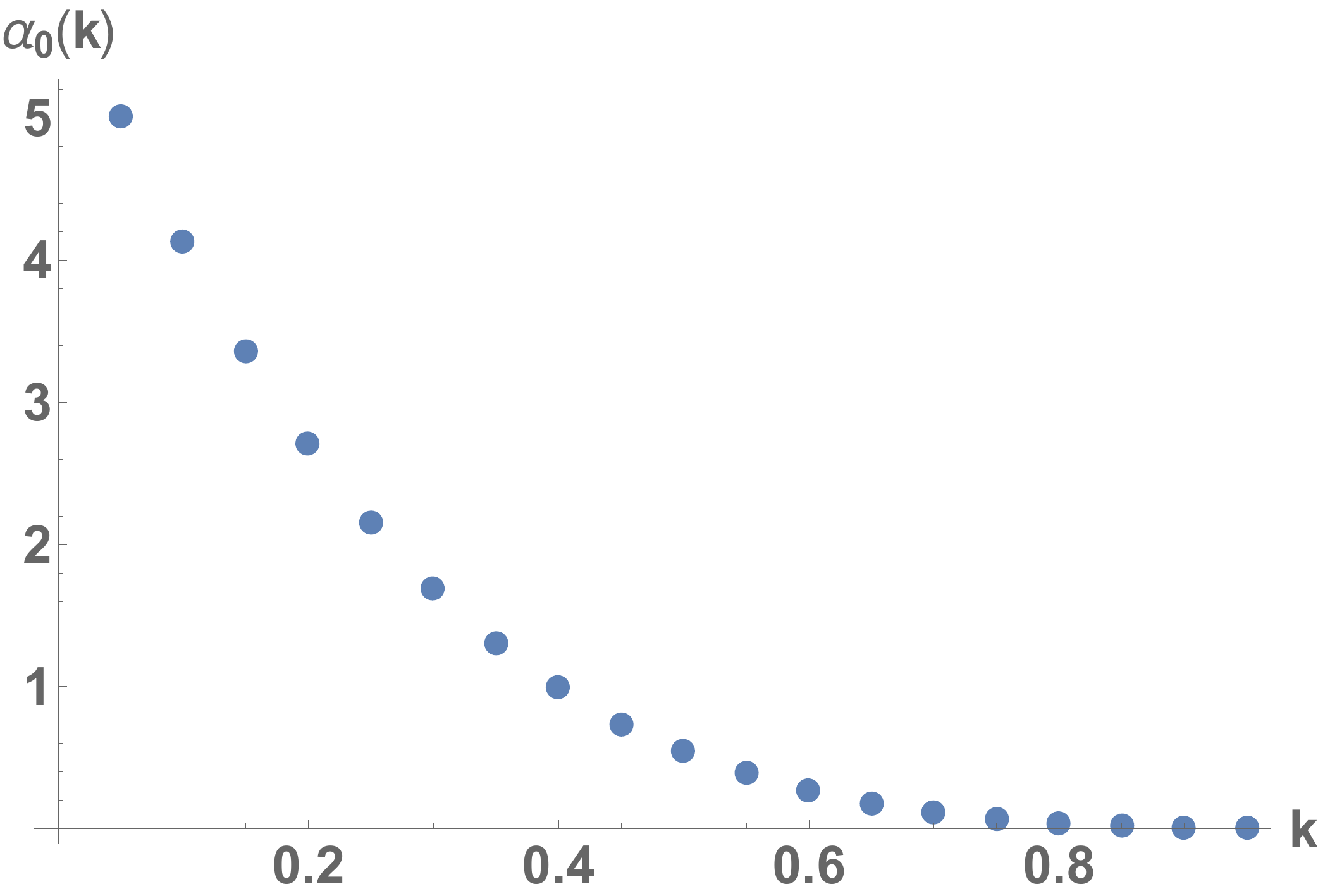}
  \caption{$\alpha_0(k)$ versus $k$.}
 \label{fig_alpha0}
 \end{center}
\end{figure}

\begin{figure}[h!]
\begin{center}
 \includegraphics[width=8cm]{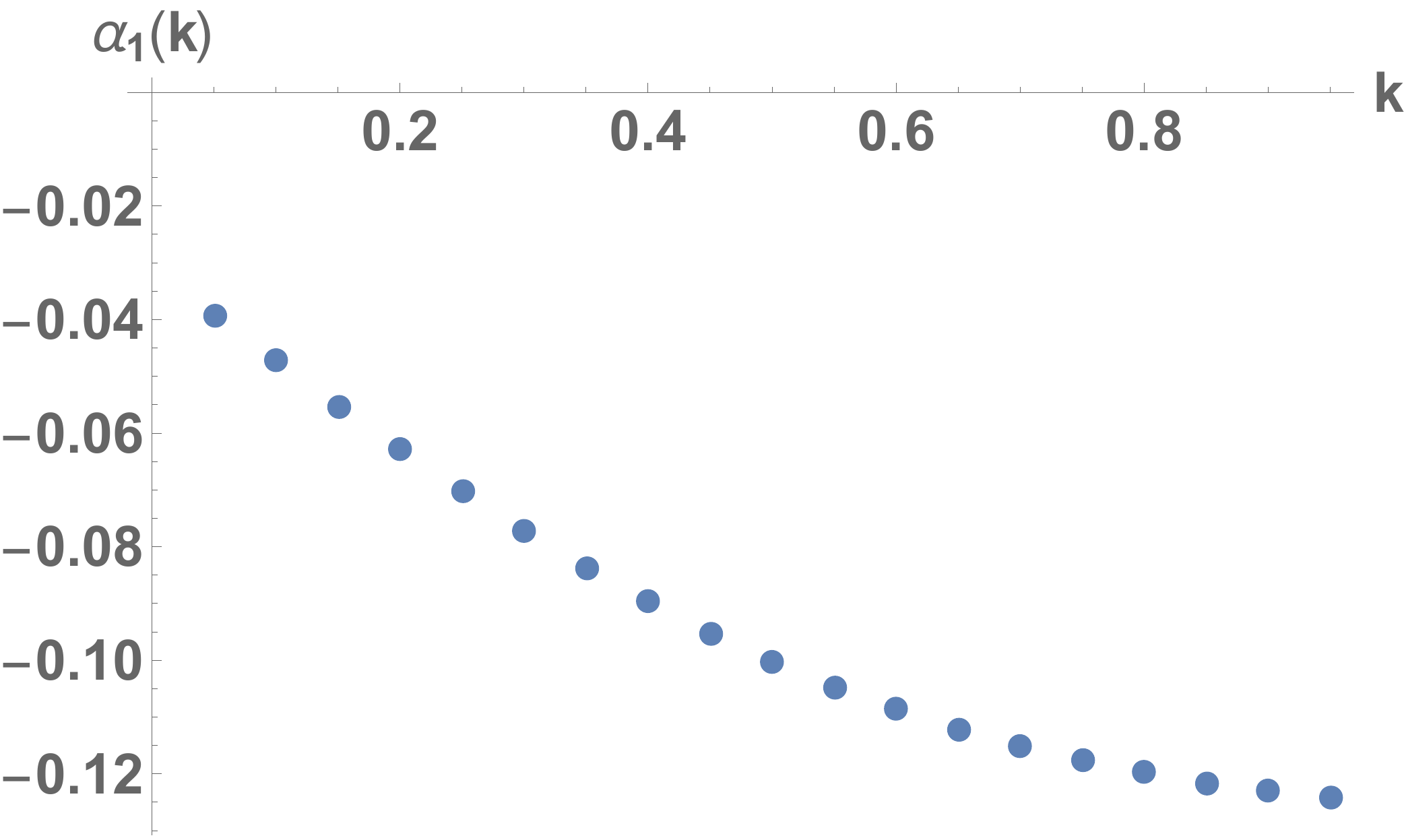}
  \caption{$\alpha_1(k)$ versus $k$.}
 \label{fig_alpha1}
 \end{center}
\end{figure}

\begin{figure}[h!]
\begin{center}
 \includegraphics[width=8cm]{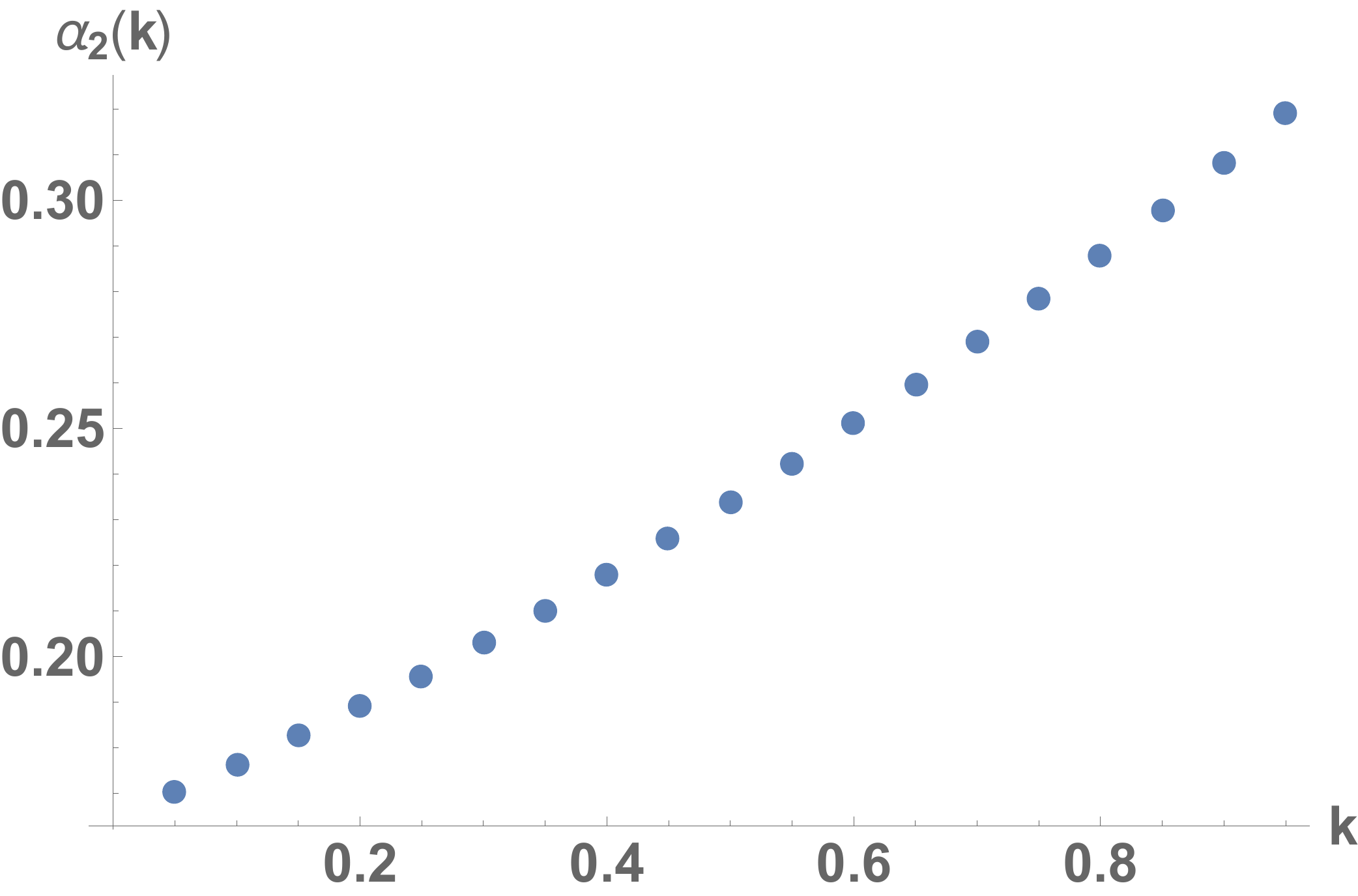}
  \caption{$\alpha_2(k)$ versus $k$.}
 \label{fig_alpha2}
 \end{center}
\end{figure}

The behavior of $\alpha_{0,1,2}(k)$ versus $k$ is shown in Figs.~\ref{fig_alpha0}-\ref{fig_alpha2}  respectively.
Eq.~$\alpha_0(k)$  is maximally repulsive at $k=0$ reflecting on the maximal deviation from self-duality, and vanishes
at  the self-dual point with $k=1$ in agreement with~\cite{LIUHEAVY}.
Eq.~$\alpha_1(k)$ is attractive for all $k$ with the expected value $\alpha_1(1)=-\frac 18$ from~\cite{LIUHEAVY}.
Eq,~$\alpha_2(k)$ is repulsive throughout with the limiting value $\alpha_2(1)=\frac 13$ in agreement with~\cite{LIUHEAVY}.

\subsection{Classical binding energy}


In general the treatment of the size $\rho$ requires quantum mechanics as we detail below. However,
for a classical estimate,
we note that since the dependence of $M_k$ versus $k$ is mild as seen in Fig.~\ref{fig_SkS1}
with a size of order $(\lambda N_c)^0/\sqrt{\lambda}$, we may
fix it near the instanton point at $k\sim 1$ with the result~\cite{SSXB,SSXBB}

\be
\label{RHO}
\rho^2=\frac{27\pi}\lambda \,\sqrt{\frac 65}
\ee
in units of $M_{KK}$.
If $N_Q=\chi_Q^\dagger \chi_Q$ is  the number of heavy quarks (mesons) attached to the tunneling configuration,
their classical binding energy $\Delta_H (k)$ as a function of $k$  follows from (\ref{SZ4}) as ($M_1=8\pi^2\kappa$)

\be
\label{BINDING}
\Delta_H(k)\equiv &&(\mathbb H_k-N_Qm_H)\nonumber\\
\equiv &&M_k+\Bigg(\frac{\lambda\alpha_0(k)}{16m_H}+162\pi \alpha_1(k)\Bigg)\frac{N_Q}{\rho^2}\nonumber\\
&&+27\pi \alpha_2(k)\,\frac{N_Q^2}{N_c\rho^2}
\ee

\begin{figure}[h!]
\begin{center}
\includegraphics[width=8cm]{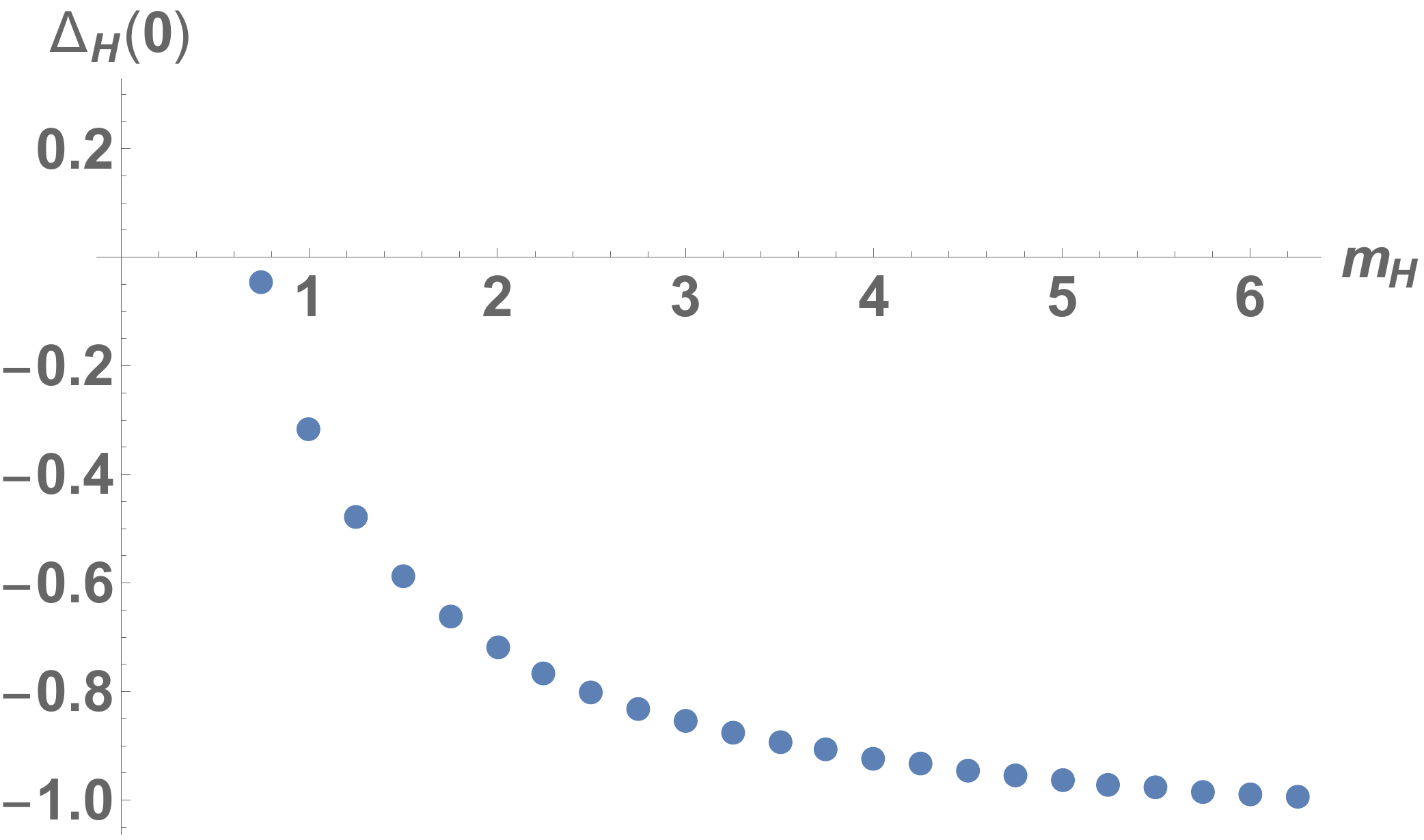}
  \caption{Classical binding energy $\Delta_{H} (0)$   versus $m_H$ for $N_Q=2$ and $\lambda=20$ in units of $M_{KK}=1$ GeV}
 \label{fig_bindh}
 \end{center}
\end{figure}

We fix the holographic parameters to:  $N_c=3$, $\lambda=20$~\cite{SSXB,ANTON}
and $M_{KK}=m_\rho/\sqrt{0.67}\approx 1\,{\rm GeV}$~\cite{SSXB}.
 For $N_Q=2$, we show in Fig.~\ref{fig_bindh} the classical binding energy $\Delta_H(0)$ 
versus $m_H$  in units of $M_{KK}$. We recall that $k=0$ corresponds to the sphaleron path
with zero topological charge. For charm and bottom, $m_H$
is  fixed  to the  $(0^-, 1^-)$ multiplet, i.e. $m_D\approx 1.870$ GeV for $(D,D^*)$  and 
 $m_B\approx 5.279$ GeV for $(B, B^*)$.  The classical binding of both charm and bottom is large and
depends sensitively on the value of $\rho$.  In the heavy quark limit
with $m_H\rightarrow\infty$, the classical binding disappears when the repulsion
($\alpha_2$) exceeds the attraction ($\alpha_1$) modulo $M_k$, i.e.
$N_Q/N_c> 6|\alpha_1(0)|/\alpha_2(0)\approx 1.22$.
A more accurate estimate of the binding
requires a quantum treatment as we now discuss.

\subsection{Quantum moduli}

\begin{figure}[h!]
\begin{center}
\includegraphics[width=8cm]{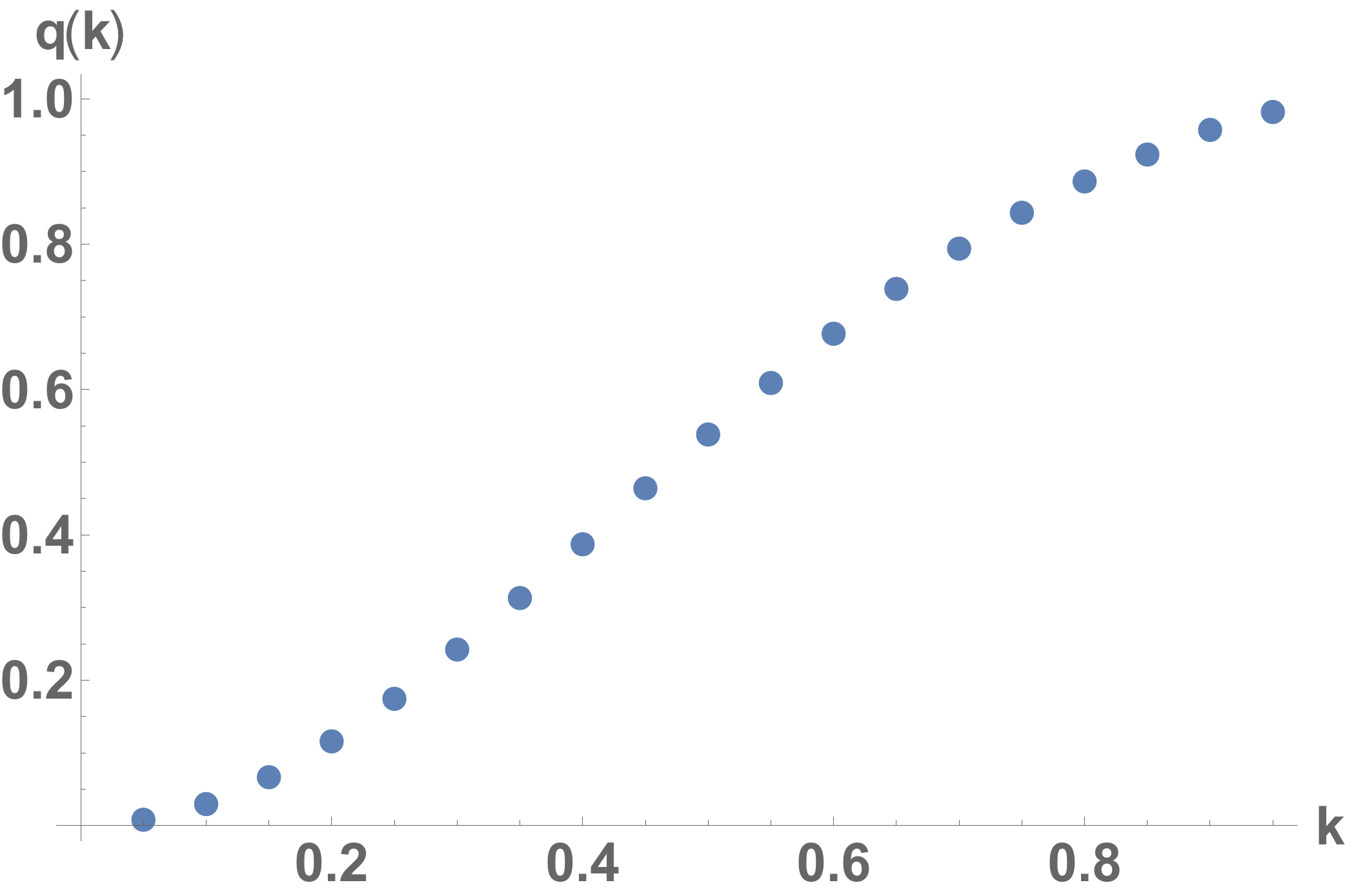}
  \caption{U(1) self-repulsion $q(k)$ versus $k$.}
 \label{fig_qk}
 \end{center}
\end{figure}

The quantum moduli space of the O(4) tunneling configurations is analogous to the instanton moduli or
$R^4\times R^4/Z_2$ (flat space)~\cite{SSX}. Here we focus on $R^4/Z_2$ which corresponds to the size
and global flavor SU(2) orientations.We will refer to $y_I=\rho a_I$ as the
coordinates on $R^4/Z_2$, with the SU(2) orientations parametrized by $a_I$ subject to the
normalization $a_I^2=1$,  and  to $\rho$ as the size of the instanton.  The collective Hamiltonian
on the $R^4/Z_2$ moduli for the bound molecule follows from the arguments given in~\cite{LIUHEAVY} as

\be
\label{H1}
{\bf H}=&&-\frac 1{2m_k}\left(\frac 1{\rho^{\frac 32}}\partial^2_\rho\,\rho^{\frac 32}
+\frac 1{\rho^2}(\nabla^2_{S^3} -2m_kQ(k))\right)\nonumber\\
&&+\frac 12 m_k\omega_k^2\rho^2
\ee
with $m_k/m_1=M_k/M_1$, 

\be
\label{H2}
&&Q(k)=\frac {N_c}{40\pi^2 a}\nonumber\\
&&\times \left(q(k)+\frac {\lambda}{m_H}\frac {5\alpha_0(k)}{432\pi}\frac {N_Q}{N_c}
+30\alpha_1(k)\frac{N_Q}{N_c}+5\alpha_2(k)\frac{N_Q^2}{N_c^2}\right)\nonumber\\
\ee
 and the inertial parameters $m_1=16\pi^2 aN_c$, $\omega_1^2=\frac 16$.
Here $q(k)$ is the U(1) topological  self-repulsion in the absence of the heavy mesons

\be
\label{H2X}
q(k)=-\frac{45}2 \,\int_{-\frac{T_k}2}^{\frac{T_k}2}\,e^{-2\xi}\phi_1(\xi)\,\mathbb F_0[\xi, k]
\ee
In Fig~\ref{fig_qk} we show  (\ref{H2X})  versus $k$, which is seen to increase monotonously
with the topological  charge, from  $q(0)=0$ on the sphaleron path  to  $q(1)=1$ on the
instanton path~\cite{LIUHEAVY}.

All contributions in (\ref{H2}) are in principle leading in the triple limit $N_c>\lambda>m_H\gg 1$ provided that
$N_Q/N_c$ is of order 1. However, in practice some of the the inequalities may  not  be fulfilled. This is a known
shortcoming of the holographic construction, where for instance $\lambda>N_c$  is used in  most applications
\cite{SSX,SSXB,ANTON}.

For $N_Q=0$, the eigenstates of (\ref{H1}) are given by $T_l(a)R_{ln}$, where $T_l(a)$ are
the spherical harmonics on $S^3$ with $\nabla^2T_l=-l(l+2)T_l$.
Under SO(4)$\sim$ SU(2)xSU(2) they are in the $(\frac l2, \frac l2)$ representations, with the
two SU(2) identified by the isometry $a_I\rightarrow V_La_IV_R$. The left factor is the isospin rotation
and the right factor is the space rotation with quantum numbers $I=J=\frac l2$.

For $N_Q\neq 0$, the isospin (${\bf I}$) and the spin  (${\bf J}$) decouple with the identification
\cite{LIUHEAVY}

\be
\label{H3}
{\bf J}=-{\bf I}+\chi^\dagger_Q{\bf T}\chi_Q
\ee
The isospin-spin representations are now

\be
\label{H4}
IJ\equiv \left(\frac l2, \frac l2\right)\rightarrow \left(\frac l2, \frac l2 \bigoplus_{i=1}^{N_Q} \frac 12\right)
\ee

\subsection{Multiquark exotics}

The radial equation for the reduced wavefunction $R_{nl}=u_{nl}/\rho^{\frac 32}$ following from (\ref{H1}) now reads

\be
\label{H5}
-u_{nl}^{\prime\prime}+\frac {g_l(k)}{\rho^2}\,u_{nl}+(m_k\omega_k\rho)^2\,u_{nl}=e_{k, nl}\,u_{nl}
\ee
with

\be
\label{H6}
g_l(k)=l(l+2)+2m_kQ(k)
\ee
and $e_{k, nl}=2m_k(E_{k, nl}-M_k-N_Qm_H)$. The quantum corrected classical
binding energy (\ref{BINDING}) is now

\be
\label{QBINDING}
\Delta_H(k)\rightarrow \Delta_k=E_{k, nl}-N_Qm_H =M_k+\frac{e_{k, nl}}{2m_k}
\ee
The occurence of the $1/\rho^2$ potential at short distances, stems from the nature of the attraction
in (\ref{SZ4}) which is dipole-like and the repulsion in (\ref{SZ4}) which is U(1)-like and Coulombic in 4-spatial
dimensions. It is dominant at small distances, with the critical coupling of $-\frac 14$ for the formation of deep
bound states below the heavy meson threshold. {\it Throughout, binding means that } $E_{k, nl}-N_Qm_H<0$.

\begin{figure}[h!]
\begin{center}
\includegraphics[width=8cm]{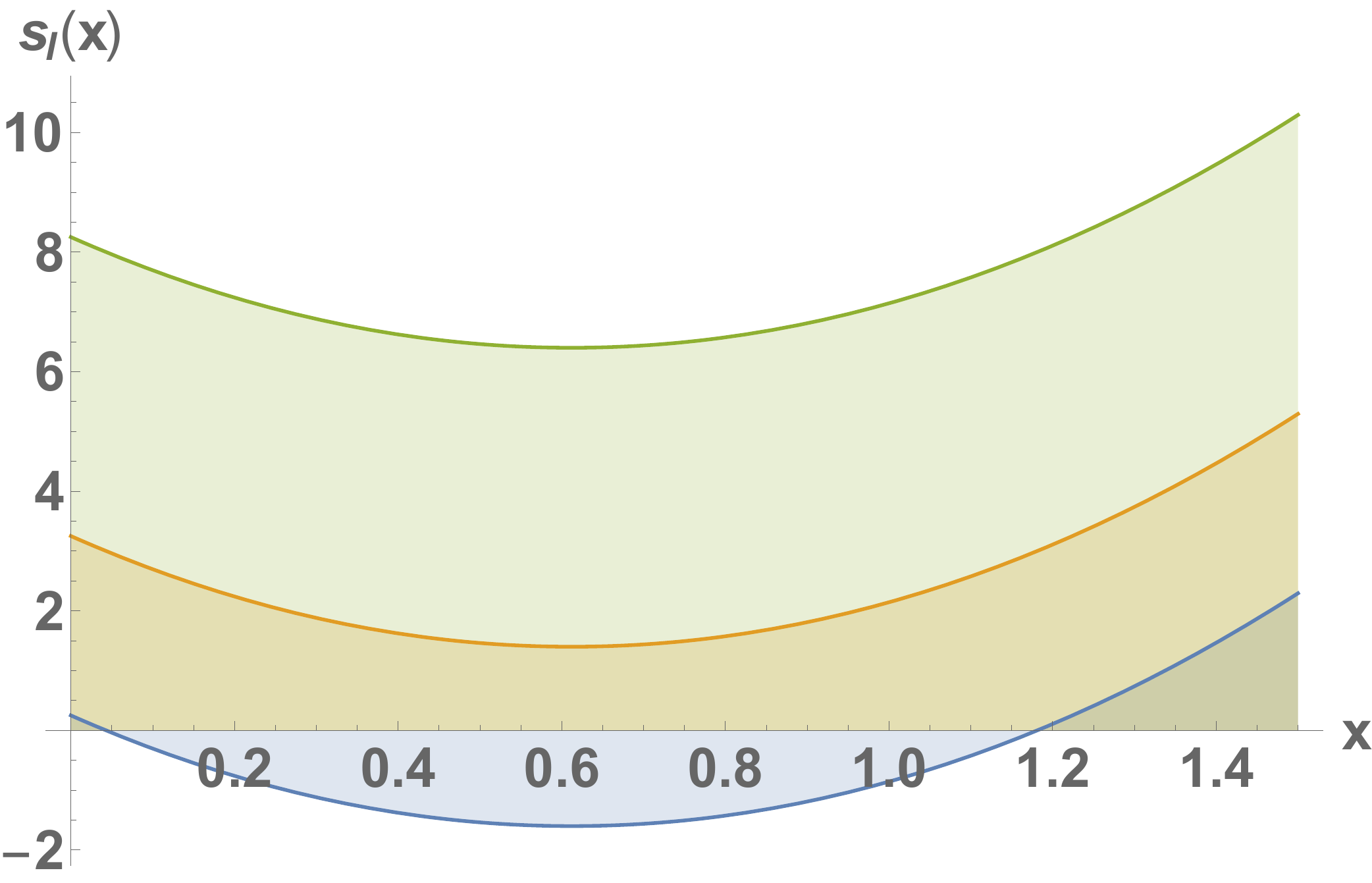}
  \caption{$s_l(x)\equiv g_l(0)+\frac 14$ versus $x={N_Q}/{N_c}$ for $l=0$ lower (blue) curve, $l=1$ middle (orange) curve and
  $l=2$ upper (green) curve.}
 \label{fig_sgn}
 \end{center}
\end{figure}

In Fig.~\ref{fig_sgn} we show the behavior of $s_l(x)\equiv g_l(0)+\frac 14$ as a function of $x=N_Q/N_c$ with  $m_H\rightarrow\infty$, on
the sphaleron  path (zero topological charge), for $l=0$ lower (blue) curve, $l=1$ middle (orange) curve and
$l=2$ upper (green) curve. Only  for $l=0$ and  $x=N_Q/N_c<1.2$ is the attraction sufficiently strong
to form deep bound states.  Higher waves with $l=1,2, ...$ are unbound. For $N_c=3$, only
$N_Q\leq 3$ states are {\it a priori} bound, i.e. open-flavor  tetraquark $QQ\bar q\bar q$ and hexaquark $QQQ\bar q\bar q\bar q$ states.
The S-wave tetraquark  states $QQ\bar q\bar q$ carry
$IJ^\pi=00^+, 01^+$ assignments  with Chern-Simons number $+\frac 12$, and are degenerate.

\subsection{Efimov states}

For small distances and S-waves, (\ref{H5}) reduces to

\be
\label{H7}
-u_{n0}^{\prime\prime}+\frac {g_0(k)}{\rho^2}\,u_{n0} \approx e_{k, n0}\,u_{n0}
\ee
For $g_0(k)+\frac 14<0$, the potential in (\ref{H7})  is sufficiently attractive to form deep bound states. However,
it is singular and requires regularization~\cite{RG}. 
The scale invariance of (\ref{H7}) allows for a universal regularization using the 
renormalization group approach, whereby the depth of the attractive and singular  potential,
can be chosen to be a function of a short distance cutoff $R_S$~\cite{RG}.
As a result, (\ref{H7}) admits many S-wave bound states with an accumulation near
threshold, the so-called Efimov states~\cite{EFIMOV}. We note that in our case, there is a  minimal value
 for the cutoff $R_{\rm min}=1/m_H$,  so the number of bound states is  limited.

\begin{figure}[h!]
\begin{center}
\includegraphics[width=8cm]{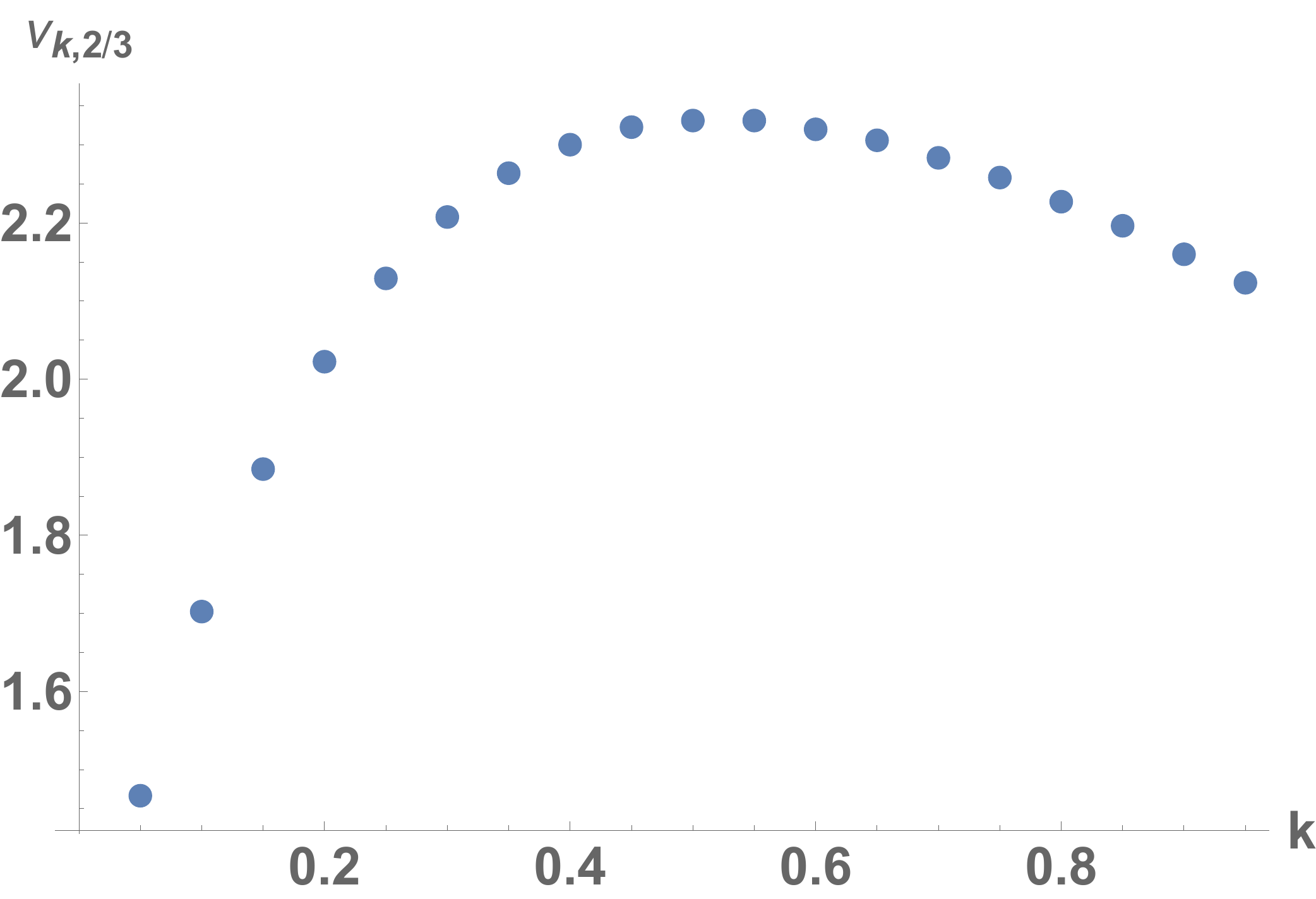}
  \caption{$\nu_{k,2/3}$ versus $k$ for $x=\frac 23$  and $m_H\rightarrow\infty$.}
 \label{fig_nuk}
 \end{center}
\end{figure}

The bound state spectrum for (\ref{H7}) was detailed in~\cite{RG} with extensive analysis in the
context of the renormalization group cycle, with the result

\be
\label{H9}
&&e_{k, n0}=-\frac 4{r_0^2}\,{e^{\varphi_{k,n}}}\nonumber\\
&&\varphi_{k,n}=\left(\frac 2{\nu_{k,x} }\left(C+{\rm Im\,ln}\,\Gamma(1+i\nu_{k,x})-\left(n+\frac 12\right)\pi\right)\right)\nonumber\\
&&\nu_{k,x}=\left(-\frac 14 -g_0(k)\right)^{\frac 12}
\ee
where $x=N_Q/N_c$. To avoid cluttering the notations, we omitted the x-dependence
in all quantities except $\nu_{k,x}$. 
 The behavior of $\nu_{k,x}$ versus $k$ is shown in Fig.~\ref{fig_nuk}
 for $N_c=3, x=2/3$ and $m_H\rightarrow\infty$. The coupling  peaks at $k=\frac 12$,
 halfway between the instanton and sphaleron path.
 Consecutive bound state energies are tied geometrically

\be
\label{H10}
\frac{e_{k, (n+1)0}}{e_{k,n0}}=e^{-\frac {2\pi}{\nu_{k,x}}}
\ee
showing  their accumulation or dissipation at threshold $e_{k,00}=0$, i.e. the  Efimov effect.
The undetermined constant $C$ in the quantum spectrum (\ref{H9}) reflects on the singular potential,
that warrants renormalization. Changing $C$ amounts to redefining the
depth of the singular potential.

In (\ref{H9}) the scale $r_0$ is fixed by the curvature or range of the long distance $1/\rho^2$ potential. From (\ref{H5}) we fix it
by matching the strength of the short distance potential to the strength of the large distance harmonic  potential, i.e.
$r_0^2= 1/(m_k\omega_1)$. As a result,  the quantum binding energy (\ref{QBINDING})  becomes

\be
\label{H11}
\Delta_k(N_c, x, \lambda)=&&M_k-2\omega_1\, {e^{\varphi_{k,n}}}\nonumber\\
=&&\left(\frac{\lambda N_c}{27\pi}\right)\,\frac{M_k}{M_1}-\sqrt{\frac 23}\, {e^{\varphi_{k,n}}}
\ee
For $N_c=3$ and at the sphaleron point with $k=0$, the binding energy (\ref{H11}) depends on the
coupling $\lambda$, the occupation ratio $x=N_Q/N_c<1.2$ and the parameter $C$ (cutoff depth). We
fix $C$ so that the binding energy vanishes for $x=1/3$ giving a  $Q\bar q$ meson state of energy exactly $m_H$.
With this in mind, (\ref{H11}) reads

\be
\label{H11X}
\Delta_0(3,\lambda, x)=&&\frac \lambda{24\sqrt{2}}
 -\sqrt{\frac 23}\,\left(\frac {\lambda\sqrt{3}}{48}\right)^{{\nu_{0,1/3}}/{\nu_{0,x}}}\nonumber\\
&&\times e^{\frac 2{\nu_{0,x}}\left({\rm Im \,Ln}\,\Gamma(1+i\nu_{0,x})-{\rm Im \,Ln}\,\Gamma(1+i\nu_{0,1/3})\right)}\nonumber\\
\ee
In the heavy quark limit, 
 $\nu_{0,1/3}=1.10$, $\nu_{0,2/3}=1.26$ and $\nu_{0, 3/3}=0.93$.

\begin{figure}[h!]
\begin{center}
\includegraphics[width=8cm]{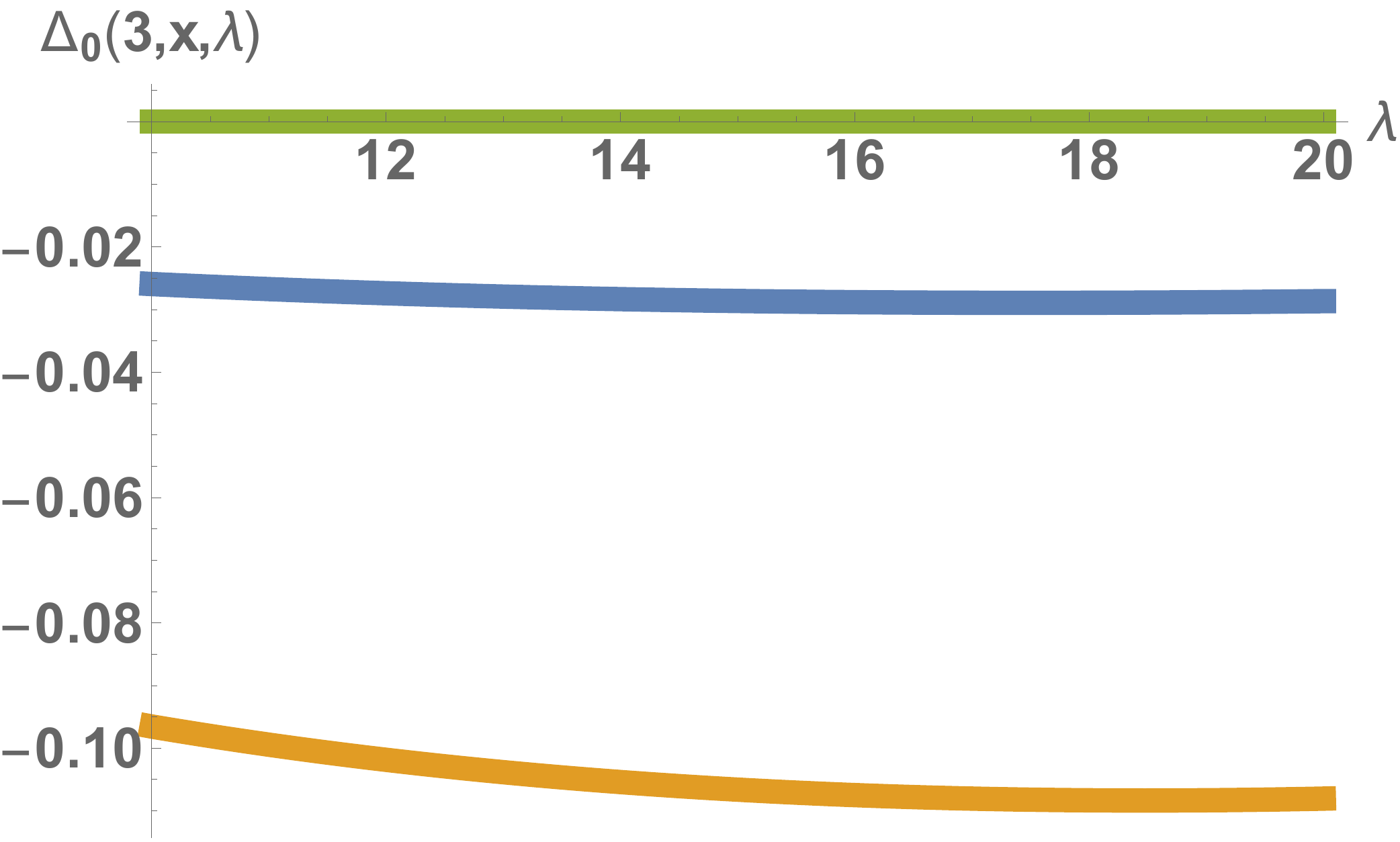}
  \caption{Binding energy (\ref{H11X}) versus $\lambda$ for $m_H\rightarrow\infty$: 
  $x=\frac 13$ green (upper) curve, $x=\frac 23$ lower (orange)  curve and $x=\frac 33$ middle (blue) curve.}
 \label{fig_dlx}
 \end{center}
\end{figure}

 Fig.~\ref{fig_dlx} shows the behavior of the binding energies (\ref{H11X}) versus the t$^\prime$ Hooft coupling $\lambda$,  for 
$x=\frac 13$ upper  (green) curve, $x=\frac 23$ lower (orange) curve and $x=\frac 33$ middle (blue) curve, 
in the heavy quark limit. For $m_H\rightarrow \infty$, the binding energies of the tetraquark states $QQ\bar q\bar q$ and hexaquark states
$QQQ\bar q\bar q\bar q$ 
 are listed  in the first column of Table~\ref{tab_bindtet}-\ref{tab_bindhex}  respectively. The binding energies are not very 
sensitive to $10\leq \lambda\leq 20$. For bottom B-mesons with $m_B=5.279$ GeV and charmed 
D-mesons with $m_D=1.87$ GeV,  the binding energies for tetraquark and hexaquark states are also
listed in Table~\ref{tab_bindtet}-\ref{tab_bindhex} respectively. The hexaquark states are unbound for finite masses.

We note that for the bound tetraquark states, the Efimov factor is 
$e^{-2\pi/\nu_{0,2/3}}\approx 10^{-3}$. This factor in the geometrical ratio (\ref{H10}) shows that the 
radially excited tetraquark exotics rapidly move to the continuum and unbind.  So we expect only one 
bound state to survive.

\begin{table}[h]
\caption{Binding energies for tetraquarks}
\begin{center}
\begin{tabular}{|c|c|c|c|c|}
\hline
$\lambda$ & $QQ\bar q\bar q$ GeV  &  $bb\bar q\bar q $ GeV  & $bc\bar q\bar q $ GeV & $cc\bar q\bar q $ GeV  \\
\hline
10 &$-0.097$ & $-0.088$   & $-0.080$&  $-0.072$\\
15 & $-0.107$  & $-0.091$ &  $-0.077 $&$ -0.062$\\
20&  $-0.108$& $-0.085$  &$-0.064$ &$ -0.041$ \\
\hline
\end{tabular}
\end{center}
\label{tab_bindtet}
\end{table}%

\begin{table}[h]
\caption{Binding energies for hexaquarks}
\begin{center}
\begin{tabular}{|c|c|c|c|c|}
\hline
$\lambda$ & $QQQ\bar q\bar q\bar q$ GeV  &  $bbb\bar q\bar q \bar q$ GeV  & $bbc\bar q\bar q \bar q$ GeV & $ccc\bar q\bar q \bar q$ GeV  \\
\hline
10 &$-0.026$& $ -0.005$ & +0.008 &  +0.037\\
15 & $-0.029$ & +0.006 &  +0.029 & +0.076\\
20&  $-0.029$&  +0.018& +0.045 & +0.084 \\
\hline
\end{tabular}
\end{center}
\label{tab_bindhex}
\end{table}%

The heavy  multiquark states with  hidden-flavor are
also covered by the present analysis provided that $N_Q\rightarrow N_Q-N_{\bar Q}$
is substituted in (\ref{H2}) when $N_Q>N_{\bar Q}$. For $N_Q<N_{\bar Q}$ the starting
O(4) tunneling configuration should be an anti-instanton path with topological charge $-1$,
interpolating  to an anti-sphaleron  path with Chern-Simons number $-\frac 12$. This
implies  conjugate symmetry for bosonic multiquark states. As a result, the
conjugate tetraquark states $\bar Q\bar Q qq$  are  bound and degenerate with $QQ\bar q\bar q$.
 Heavier  tetraquark states such as
$QQ\bar Q\bar Q$ with $N_Q\rightarrow N_Q-N_{\bar Q}=0$
are  unbound.  Heavier multiquark states  of the type $QQ\bar Q\bar q$ and 
$QQ\bar Q q\bar q\bar q$ with $N_Q\rightarrow N_Q-N_{\bar Q}=1$ are  unbound, but those
of the type $QQ(Q\bar Q)^p\bar q\bar q$ with $N_Q-N_{\bar Q}=(p+2)-p=2$ are  bound in the heavy quark limit.

The holographic exotics with hidden-flavor $Q\bar Qq\bar q$ with $N_Q\rightarrow N_Q-N_{\bar Q}=0$,
are also unbound. Experimental evidence in~\cite{BELLE,BESIII,DO,LHCb} suggests otherwise.  This
shows  that the X,Y,Z states reported in~\cite{BELLE,BESIII,DO,LHCb} are not the compact topological
molecules discussed here, but  likely loosely bound hadronic molecules (deusons)~\cite{MOLECULES,THORSSON,KARLINER,OTHERS,OTHERSX,OTHERSZ,OTHERSXX,LIUMOLECULE}.
Finally, we also note that the present analysis is limited to  the  light SU(2) flavor sector. The extension to the SU(3) flavor
sector with massive strange quarks  is more involved~\cite{LIUHEAVY} and will be discussed elsewhere.

\subsection{Discussion}

Recent  lattice and phenomenological estimates suggest that the double-bottom tetraquark state is deeply
bound with $\Delta_{BB}=-(0.15-0.2)$ GeV~\cite{MALT} (lattice) and $\Delta_{BB}=-(0.17)$ GeV~\cite{KR}
(and references therein). The same lattice analysis suggests that the mixed charm-bottom tetraquark state is bound
$\Delta_{CB}=-(0.061-0.015)$ GeV, but the the double-charm tetraquark state is not~\cite{MALT}.

Our holographic
results support a double-bottom tetraquark state with a binding  energy  $\Delta_{BB}=-(0.088-0.091)$ GeV
somewhat lower than the lattice estimate,  a mixed bottom-charm tetraquark with a binding  energy  
$\Delta_{CB}=-(0.064-0.080)$ GeV closer to the lattice estimate, and a {\it bound} 
double-charm tetraquark with a binding  energy  $\Delta_{CC}=-(0.041-0.072)$ GeV contrary to
the lattice estimate.  These results are overall consistent with earlier estimates in the context of the random
instanton model~\cite{MACIEK3}.

Finally, we note that multiquark exotics in the context of holography have been recently
addressed in the context of the holography inspired stringy hadron model (HISH)~\cite{COBIX}.  We view our
analysis as complementary to the HISH analysis as it applies to the low lying exotics as opposed
to the highly excited and stringy exotics. Our analysis can be extended to the excited and unbound exotic states
for slow rotations, with the rotational-vibrational spectrum for $l>0$

\be
\label{H14}
E_{0,nl}=&&M_0+N_Qm_H\nonumber\\
&&+\left(\frac{l^2}6+\frac {2N_c}{81}Q(0)\right)^{\frac 12}+\frac{2n+1}{\sqrt{6}}\nonumber\\
\ee
following from (\ref{H1}-\ref{H2}) through standard arguments~\cite{SSX,LIUHEAVY}.
The  states described by (\ref{H4}) are unstable against the strong decay   to heavy-light mesons.
For $l\gg 1$ the spectrum (\ref{H14}) does not reggeize since $E_{0,nl}\approx l$. A way to
achieve reggeization  is through {\it relativistic} rotations, that allow
for a stringy-like deformation of the underlying O(4) tunneling configuration.   This will
be discussed elsewhere.

\section{Conclusions}

We have presented a top-down holographic approach to multiquark exotic states using  a
minimally modified D4-D8-D$\bar 8$ set up to account for two light and one heavy flavor~\cite{LIUHEAVY}.
The heavy multiquark states are topological molecules of heavy-light mesons bound to a tunneling
 gauge configuration  with fixed Chern-Simons number. The latter interpolates between an instanton path  with
 net topological charge 1 or baryon number (a fermion),  and a sphaleron path with net topological charge 0 and
Chern-Simons number $\frac 12$ (a boson).

The geometrical interpolation between a fermion and a boson in higher dimensions, is remarkable. It  
points to a topological duality between the heavy baryon  exotics discussed in~\cite{LIUHEAVY},  and the heavy  
meson exotics addressed here. This is suggestive of a geometrical realization and generalization of 
the Savage-Wise symmetry~\cite{SW} to most  heavy exotics.

In leading order in the heavy quark limit, the bounded heavy mesons to the tunneling path with fixed Chern-Simons number
transmute to fermions. This mechanism  
is reminiscent  of the  transmutation of the strange quark spin to the Skyrmion in the kaon-Skyrmion bound state
~\cite{CALLANKLEB}.  The binding of the fermions follows by
balancing the attraction induced by the Chern Simons term which is dipole-like, and the dual repulsion stemming
from the induced U(1) gauge field together with the deviation of the tunneling configuration from
self-duality which are 4-dimensional Coulomb-like. The ensuing potential in the molecule is singular.  As a result the topologically
bound exotics are Efimov states.

Our analysis shows that only the open-flavor molecules with  $x=N_Q/N_c<1.2$ are bound in the heavy quark limit.
For $N_c=3$, the open-flavor exotics $QQ\bar q \bar q$ and their conjugate $\bar Q\bar Q q q$ are bound in {\it a degenerate multiplet}
$IJ^\pi=(00^+ , 01^+)$ with opposite intrinsic Chern-Simons numbers $\pm \frac 12$.  
The open-flavor and non-strange hexaquark states $QQQ\bar q\bar q \bar q$ are bound 
in the heavy quark limit only. The 
heavier exotics $QQ\bar Q\bar Q$  are unbound. The compact exotics with hidden-flavor such as $QQ\bar Q\bar q$ and
 $Q\bar Q q\bar q$ are also unbound, but the heavier exotics such as $QQ(Q\bar Q)\bar q\bar q$ are  bound in the heavy quark limit.

The leading holographic correction in the heavy quark mass
is found to penalize the binding in $cc\bar q\bar q$ more than in $bb\bar q\bar q$. Our analysis suggests a rotational-vibrational 
tower of multiquark excitations prone to strong decay.

Some of the shortcomings of the present approach lie  in the use of the triple limits of large $N_c$,  large
coupling $\lambda$,  and  large meson mass $m_H$. Although the relaxation of these limits is
straightforward in principle, its systematic implementation  is involved in practice. This not withstanding,
the present setup  is noticeable  because of  the limited number of parameters it carries.
The brane tension $\kappa\sim \lambda N_c$ is usually traded for the pion decay constant,  and the
KK scale for  the rho meson mass all in the light meson sector, leaving the treatment of
the heavy-light sector parameter free modulo the heavy meson masses $m_H$.

Unlike most of the approaches for heavy exotics~\cite{NEW} (and references therein), the present construction enforces heavy quark symmetry, 
the spontaneous breaking of chiral symmetry, provides a systematic organizational framework using the QCD parameters
$N_c, \lambda, m_H$, and solves
the multi-body problem using topological bound states. 

The construction can be improved in a number of ways. For instance,  by breaking isospin symmetry and
including strangeness
to account for strange topological exotics, or by adding a tachyon and a tachyon potential to bring the
model closer to QCD at short distances, perhaps in
the context of improved holographic QCD~\cite{KIRITSIS}. These and related issues will be discussed next.

\section{Acknowledgements}

We thank Marek Karliner for a discussion.
MAN thanks the Nuclear Theory Group  at  Stony Brook University for 
hospitality during the completion of this work.
This work was supported by the U.S. Department of Energy under Contract No.
DE-FG-88ER40388 and by the Polish National  Science Centre (NCN) Grant 
UMO-2017/27/B/ST2/01139.

\section{Appendix I: Heavy-light action}


The explicit construction of the holographic heavy-ligh  action was detailed in~\cite{LIUHEAVY}.
Here we quote the relevant expressions for (\ref{RS1}) for completeness,

\begin{eqnarray}
\label{RX0}
{\cal L}_0=&&\nonumber -(D_{M}\Phi_{N}^{\dagger}-D_{N}\Phi_{M}^{\dagger})(D_M\Phi_N-D_N\Phi_M)\nonumber \\
&&+2\Phi_{M}^{\dagger}F_{MN}\Phi_{N}\nonumber\\
{\cal L}_1=&&+2(D_{0}\Phi_{M}^{\dagger}-D_{M}\Phi_{0}^{\dagger})(D_0\Phi_M-D_M\Phi_0)\nonumber \\
&&-2\Phi_0^{\dagger}F^{0M}\Phi_{M}-2\Phi_{M}^{\dagger}F^{M0}\Phi_0\nonumber\\
&&-2m_H^2\Phi_{M}^{\dagger}\Phi_M +\tilde {\cal L}_1\nonumber \\
{\cal L}_{CS}=&&-\frac{iN_c}{24\pi^2}(d\Phi^{\dagger}Ad\Phi+d\Phi^{\dagger}dA\Phi+\Phi^{\dagger}dAd\Phi)\nonumber \\
&&-\frac{iN_c}{16\pi^2}(d\Phi^{\dagger} A^2\Phi+\Phi^{\dagger}A^2d\Phi+\Phi^{\dagger}(AdA+dAA)\Phi)\nonumber \\
&&-\frac{5iN_c}{48\pi^2}\Phi^{\dagger}A^3\Phi+S_C(\Phi^4,A)
\end{eqnarray}
 and

\begin{eqnarray}
\tilde {\cal L}_1=&&+\frac{1}{3}z^2(D_i\Phi_j-D_j\Phi_i)^{\dagger}(D_i\Phi_j-D_j\Phi_i)\nonumber \\
&&-2z^2(D_i\Phi_z-D_z\Phi_i)^{\dagger}(D_i\Phi_z-D_z\Phi_i)\nonumber\\
&&-\frac{2}{3}z^2\Phi_i^{\dagger}F_{ij}\Phi_j+2z^2(\Phi_z^{\dagger}F_{zi}\Phi_i+{\rm c.c.})
\end{eqnarray}

\section{Appendix II: O(3) symmetric tunneling solution and its fermionic zero mode}

\begin{figure}[h!]
\begin{center}
\includegraphics[width=6cm]{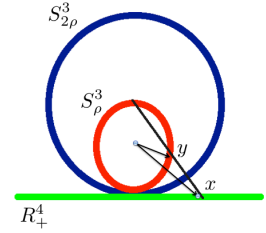}
  \caption{Inversion of $S_\rho^3$ onto $R_+^4$ through $S^3_{2\rho}$.}
 \label{fig_inverse}
 \end{center}
\end{figure}

In this Appendix we suggest yet another tunneling configuration with O(3) instead of
O(4) symmetry that is also suitable for binding heavy-light mesons which is fully localized in
flat $R^4$. This configuration is characterized by a turning point in the holographic direction
at $z=0$, in agreement with the explosive sphaleron configurations discussed in~\cite{LS,EXPLO1,EXPLO2}.
To construct it, we note that the  O(4) solutions to the Yang-Mills equations with a turning point
at $\xi=0$ relates to the solution with a turning point at $z=0$ by  the inversion

\be
(x+a)_M =\frac {2\rho^2}{|y+a|^2}\,(y+a)_M
\label{O28}
\ee
with $a=(\vec{0}, \rho)$, which maps the sphere $y^2=\rho^2$ onto
the upper-half of the x-space  as illustrated in Fig.~\ref{fig_inverse}.
This inversion leaves the line element in $R^4$ unchanged modulo
a conformal weight  $\sigma(y)$

\be
\label{O29}
|dy|^2=\sigma(y)\,|dx|^2=\frac{|y+a|^4}{4\rho^4} |dx|^2
\ee
and leaves invariant the 1-form of the gauge field

\be
\label{O30}
dx_\mu A_\mu (x)= dy_\nu A_\nu (y)
\ee
This leads to the transform

\be
\label{O31}
&&A_{aM} (x) =\nonumber\\
&&\sqrt{\sigma(y)}
\left(g_{MN}-2\,\frac{(y+a)_M(y+a)_N}{(y+a)^2}\right)\,A_{aN}(y)\nonumber\\
\ee
with $y$ solving (\ref{O28}).

We now proceed to construct the O(3) symmetric zero mode  by applying the spatial inversion (\ref{O28})
onto  the O(4) symmetric zero mode in (\ref{OZ7}) through

\be
{\tilde \psi}_+(x) = \frac{\sigma^\dagger_\mu \,(y +a)_\mu}{1/(y+a)^2}
\,{ \psi}_+ (y)
\label{x1}
\ee
More explicitly, we have ($r=|\vec x|$)

\be
{\tilde \psi}_+ (r,z) = &&
\frac{{8\rho^6\bf\, C}}{((z+\rho)^2+r^2)^2}\nonumber\\
&&\times \left((z+\rho) +i\vec{\sigma}\cdot\vec{x}\right)\,e^{-\frac 32 F(\xi(y))}\,\chi_Q\nonumber\\
\label{ZX19}
\ee
with

\be
\xi(y)=
\frac 12 {\rm ln}\left(\frac {(z-\rho)^2+r^2}{(z+\rho)^2+r^2}\right)
\ee
This result is in agreement with the one derived in~\cite{EXPLO2}  prior to the analytical
continuation to Minkowski space (see their Eq. 22  with a minor correction of the 2 to
$\frac 32$ in their exponent).

 \vfil


\newpage




\begin{thebibliography}{99} \frenchspacing







\bibitem{BELLE}
  I.~Adachi [Belle Collaboration],
  arXiv:1105.4583 [hep-ex];
  A.~Bondar {\it et al.} [Belle Collaboration],
  Phys.\ Rev.\ Lett.\  {\bf 108}, 122001 (2012)
  [arXiv:1110.2251 [hep-ex]].


\bibitem{BESIII}
  M.~Ablikim {\it et al.} [BESIII Collaboration],
  Phys.\ Rev.\ Lett.\  {\bf 110}, 252001 (2013)
  [arXiv:1303.5949 [hep-ex]].



\bibitem{DO}
  V.~M.~Abazov {\it et al.} [D0 Collaboration],
  [arXiv:1602.07588 [hep-ex]].

\bibitem{LHCb}
  R.~Aaij {\it et al.} [LHCb Collaboration],
  arXiv:1606.07895 [hep-ex];
  R.~Aaij {\it et al.} [LHCb Collaboration],
  arXiv:1606.07898 [hep-ex].


  \bibitem{LHCbx}
  R.~Aaij {\it et al.} [LHCb Collaboration],
  Phys.\ Rev.\ Lett.\  {\bf 115} (2015) 072001
  [arXiv:1507.03414 [hep-ex]];
  R.~Aaij {\it et al.} [LHCb Collaboration],
  Phys.\ Rev.\ Lett.\  {\bf 117} (2016) no.8,  082002
  [arXiv:1604.05708 [hep-ex]];
  R.~Aaij {\it et al.} [LHCb Collaboration],
  Phys.\ Rev.\ Lett.\  {\bf 117} (2016) no.8,  082003
   Addendum: [Phys.\ Rev.\ Lett.\  {\bf 117} (2016) no.10,  109902]
  [arXiv:1606.06999 [hep-ex]].


\bibitem{LHCbxx}
  R.~Aaij {\it et al.} [LHCb Collaboration],
  arXiv:1703.04639 [hep-ex].




\bibitem{MOLECULES}
  M.~B.~Voloshin and L.~B.~Okun,
  JETP Lett.\  {\bf 23}, 333 (1976)
  [Pisma Zh.\ Eksp.\ Teor.\ Fiz.\  {\bf 23}, 369 (1976)];

\bibitem{THORSSON}
  N.~A.~Tornqvist,
  Phys.\ Rev.\ Lett.\  {\bf 67}, 556 (1991);
  N.~A.~Tornqvist,
  Z.\ Phys.\ C {\bf 61}, 525 (1994)
  [hep-ph/9310247];
  N.~A.~Tornqvist,
  Phys.\ Lett.\ B {\bf 590}, 209 (2004)
  [hep-ph/0402237].







\bibitem{KARLINER}
  M.~Karliner and H.~J.~Lipkin,
  arXiv:0802.0649 [hep-ph];
  M.~Karliner and J.~L.~Rosner,
  Phys.\ Rev.\ Lett.\  {\bf 115} (2015) no.12,  122001
  [arXiv:1506.06386 [hep-ph]];
  M.~Karliner,
  Acta Phys.\ Polon.\ B {\bf 47}, 117 (2016).



\bibitem{OTHERS}
  C.~E.~Thomas and F.~E.~Close,
  Phys.\ Rev.\ D {\bf 78}, 034007 (2008)
  [arXiv:0805.3653 [hep-ph]];
  F.~Close, C.~Downum and C.~E.~Thomas,
  Phys.\ Rev.\ D {\bf 81}, 074033 (2010)
  [arXiv:1001.2553 [hep-ph]].


  \bibitem{OTHERSX}
  S.~Ohkoda, Y.~Yamaguchi, S.~Yasui, K.~Sudoh and A.~Hosaka,
  Phys.\ Rev.\ D {\bf 86}, 034019 (2012)
  [arXiv:1202.0760 [hep-ph]];
  S.~Ohkoda, Y.~Yamaguchi, S.~Yasui, K.~Sudoh and A.~Hosaka,
  arXiv:1209.0144 [hep-ph].


\bibitem{OTHERSZ}
  M.~T.~AlFiky, F.~Gabbiani and A.~A.~Petrov,
  Phys.\ Lett.\ B {\bf 640}, 238 (2006)
  [hep-ph/0506141];
  I.~W.~Lee, A.~Faessler, T.~Gutsche and V.~E.~Lyubovitskij,
  Phys.\ Rev.\ D {\bf 80}, 094005 (2009)
  [arXiv:0910.1009 [hep-ph]];
  M.~Suzuki,
  Phys.\ Rev.\ D {\bf 72}, 114013 (2005)
  [hep-ph/0508258];
  J.~R.~Zhang, M.~Zhong and M.~Q.~Huang,
  Phys.\ Lett.\ B {\bf 704}, 312 (2011)
  [arXiv:1105.5472 [hep-ph]];
  D.~V.~Bugg,
  Europhys.\ Lett.\  {\bf 96}, 11002 (2011)
  [arXiv:1105.5492 [hep-ph]];
  J.~Nieves and M.~P.~Valderrama,
  Phys.\ Rev.\ D {\bf 84}, 056015 (2011)
  [arXiv:1106.0600 [hep-ph]];
  M.~Cleven, F.~K.~Guo, C.~Hanhart and U.~G.~Meissner,
  Eur.\ Phys.\ J.\ A {\bf 47}, 120 (2011)
  [arXiv:1107.0254 [hep-ph]];
  T.~Mehen and J.~W.~Powell,
  Phys.\ Rev.\ D {\bf 84}, 114013 (2011)
  [arXiv:1109.3479 [hep-ph]];
  F.~K.~Guo, C.~Hidalgo-Duque, J.~Nieves and M.~P.~Valderrama,
  Phys.\ Rev.\ D {\bf 88}, 054007 (2013)
  [arXiv:1303.6608 [hep-ph]];
  Q.~Wang, C.~Hanhart and Q.~Zhao,
  Phys.\ Rev.\ Lett.\  {\bf 111}, no. 13, 132003 (2013)
  [arXiv:1303.6355 [hep-ph]];
  F.~K.~Guo, C.~Hanhart, Q.~Wang and Q.~Zhao,
  Phys.\ Rev.\ D {\bf 91} (2015) no.5,  051504
  [arXiv:1411.5584 [hep-ph]];
  X.~W.~Kang, Z.~H.~Guo and J.~A.~Oller,
  Phys.\ Rev.\ D {\bf 94} (2016) no.1,  014012
  [arXiv:1603.05546 [hep-ph]];
  X.~W.~Kang and J.~A.~Oller,
  arXiv:1612.08420 [hep-ph].



 \bibitem{OTHERSXX}
  E.~S.~Swanson,
  Phys.\ Rept.\  {\bf 429}, 243 (2006)
  [hep-ph/0601110];
  Z.~F.~Sun, J.~He, X.~Liu, Z.~G.~Luo and S.~L.~Zhu,
  Phys.\ Rev.\ D {\bf 84}, 054002 (2011)
  [arXiv:1106.2968 [hep-ph]];


 \bibitem{LIUMOLECULE}
  Y.~Liu and I.~Zahed,
  Phys.\ Lett.\ B {\bf 762}, 362 (2016)
  [arXiv:1608.06535 [hep-ph]];
  Y.~Liu and I.~Zahed,
  Int.\ J.\ Mod.\ Phys.\ E {\bf 26},  1740017 (2017)
  [arXiv:1610.06543 [hep-ph]];
  Y.~Liu and I.~Zahed,
  arXiv:1611.04400 [hep-ph].


\bibitem{Albaladejo:2015lob}
  M.~Albaladejo, F.~K.~Guo, C.~Hidalgo-Duque and J.~Nieves,
  Phys.\ Lett.\ B {\bf 755}, 337 (2016)
  doi:10.1016/j.physletb.2016.02.025
  [arXiv:1512.03638 [hep-ph]].





\bibitem{MANOHAR}
  A.~V.~Manohar and M.~B.~Wise,
  Nucl.\ Phys.\ B {\bf 399}, 17 (1993)
  [hep-ph/9212236];
  N.~Brambilla {\it et al.},
  Eur.\ Phys.\ J.\ C {\bf 71}, 1534 (2011)
  [arXiv:1010.5827 [hep-ph]];
  M.~B.~Voloshin,
  Prog.\ Part.\ Nucl.\ Phys.\  {\bf 61}, 455 (2008)
  [arXiv:0711.4556 [hep-ph]];
  J.~M.~Richard,
  arXiv:1606.08593 [hep-ph].


\bibitem{RISKA}
  D.~O.~Riska and N.~N.~Scoccola,
  Phys.\ Lett.\ B {\bf 299}, 338 (1993).




\bibitem{SUHONG}
  M.~Nielsen, F.~S.~Navarra and S.~H.~Lee,
  Phys.\ Rept.\  {\bf 497}, 41 (2010)
  [arXiv:0911.1958 [hep-ph]].




\bibitem{RICHARD}
  J.~P.~Ader, J.~M.~Richard and P.~Taxil,
  Phys.\ Rev.\ D {\bf 25}, 2370 (1982);
  S.~Zouzou, B.~Silvestre-Brac, C.~Gignoux and J.~M.~Richard,
  Z.\ Phys.\ C {\bf 30}, 457 (1986);
  J.~Carlson, L.~Heller and J.~A.~Tjon,
  Phys.\ Rev.\ D {\bf 37}, 744 (1988).



\bibitem{MACIEK2}
  M.~A.~Nowak, I.~Zahed and M.~Rho,
  Phys.\ Lett.\ B {\bf 303}, 130 (1993).

\bibitem{MACIEK3}
  S.~Chernyshev, M.~A.~Nowak and I.~Zahed,
  Phys.\ Rev.\ D {\bf 53}, 5176 (1996)
  [hep-ph/9510326].






\bibitem{KR}
  M.~Karliner, J.~L.~Rosner and T.~Skwarnicki,
  Ann.\ Rev.\ Nucl.\ Part.\ Sci.\  {\bf 68}, 17 (2018)
  [arXiv:1711.10626 [hep-ph]];
  M.~Karliner and J.~L.~Rosner,
  Phys.\ Rev.\ Lett.\  {\bf 119}, no. 20, 202001 (2017)
  [arXiv:1707.07666 [hep-ph]];
  E.~J.~Eichten and C.~Quigg,
  Phys.\ Rev.\ Lett.\  {\bf 119}, no. 20, 202002 (2017)
  [arXiv:1707.09575 [hep-ph]];
  A.~Czarnecki, B.~Leng and M.~B.~Voloshin,
  Phys.\ Lett.\ B {\bf 778}, 233 (2018)
  [arXiv:1708.04594 [hep-ph]].
  
  




\bibitem{MAREK}
  M.~Karliner and J.~L.~Rosner,
  Phys.\ Rev.\ Lett.\  {\bf 115}, no. 12, 122001 (2015)
  [arXiv:1506.06386 [hep-ph]];
  M.~Karliner,
  EPJ Web Conf.\  {\bf 130}, 01003 (2016).





\bibitem{MANY}
  R.~Chen, X.~Liu, X.~Q.~Li and S.~L.~Zhu,
  Phys.\ Rev.\ Lett.\  {\bf 115}, no. 13, 132002 (2015)
  [arXiv:1507.03704 [hep-ph]];
  H.~X.~Chen, W.~Chen, X.~Liu, T.~G.~Steele and S.~L.~Zhu,
  Phys.\ Rev.\ Lett.\  {\bf 115}, no. 17, 172001 (2015)
  [arXiv:1507.03717 [hep-ph]];
  L.~Roca, J.~Nieves and E.~Oset,
  Phys.\ Rev.\ D {\bf 92}, no. 9, 094003 (2015)
  [arXiv:1507.04249 [hep-ph]];
  T.~J.~Burns,
  Eur.\ Phys.\ J.\ A {\bf 51}, no. 11, 152 (2015)
  [arXiv:1509.02460 [hep-ph]].
  H.~Huang, C.~Deng, J.~Ping and F.~Wang,
  Eur.\ Phys.\ J.\ C {\bf 76}, no. 11, 624 (2016)
  [arXiv:1510.04648 [hep-ph]];
  L.~Roca and E.~Oset,
  Eur.\ Phys.\ J.\ C {\bf 76}, no. 11, 591 (2016)
  [arXiv:1602.06791 [hep-ph]];
  Q.~F.~L and Y.~B.~Dong,
  Phys.\ Rev.\ D {\bf 93}, no. 7, 074020 (2016)
  [arXiv:1603.00559 [hep-ph]];
  Y.~Shimizu, D.~Suenaga and M.~Harada,
  Phys.\ Rev.\ D {\bf 93}, no. 11, 114003 (2016)
  [arXiv:1603.02376 [hep-ph]];
  C.~W.~Shen, F.~K.~Guo, J.~J.~Xie and B.~S.~Zou,
  Nucl.\ Phys.\ A {\bf 954}, 393 (2016)
  [arXiv:1603.04672 [hep-ph]];
  M.~I.~Eides, V.~Y.~Petrov and M.~V.~Polyakov,
  Phys.\ Rev.\ D {\bf 93}, no. 5, 054039 (2016)
  [arXiv:1512.00426 [hep-ph]];
  I.~A.~Perevalova, M.~V.~Polyakov and P.~Schweitzer,
  Phys.\ Rev.\ D {\bf 94}, no. 5, 054024 (2016)
  [arXiv:1607.07008 [hep-ph]];
  V.~Kopeliovich and I.~Potashnikova,
  Phys.\ Rev.\ D {\bf 93}, no. 7, 074012 (2016);
  Y.~Yamaguchi and E.~Santopinto,
  arXiv:1606.08330 [hep-ph];
  S.~Takeuchi and M.~Takizawa,
  Phys.\ Lett.\ B {\bf 764}, 254 (2017)
  [arXiv:1608.05475 [hep-ph]].


  \bibitem{PENTARHO}
  N.~N.~Scoccola, D.~O.~Riska and M.~Rho,
  Phys.\ Rev.\ D {\bf 92}, no. 5, 051501 (2015)
  [arXiv:1508.01172 [hep-ph]].




 \bibitem{VENEZIANO}
  G.~Rossi and G.~Veneziano,
  JHEP {\bf 1606}, 041 (2016)
  [arXiv:1603.05830 [hep-th]].


\bibitem{COBI}
  J.~Sonnenschein and D.~Weissman,
  arXiv:1606.02732 [hep-ph].










\bibitem{OMEGAC}
  M.~Karliner and J.~L.~Rosner,
  arXiv:1703.07774 [hep-ph];
  G.~Yang and J.~Ping,
  arXiv:1703.08845 [hep-ph];
  K.~L.~Wang, L.~Y.~Xiao, X.~H.~Zhong and Q.~Zhao,
  arXiv:1703.09130 [hep-ph];
  W.~Wang and R.~L.~Zhu,
  arXiv:1704.00179 [hep-ph];
  H.~Y.~Cheng and C.~W.~Chiang,
  arXiv:1704.00396 [hep-ph];
  H.~Huang, J.~Ping and F.~Wang,
  arXiv:1704.01421 [hep-ph];
  B.~Chen and X.~Liu,
  arXiv:1704.02583 [hep-ph];
  T.~M.~Aliev, S.~Bilmis and M.~Savci,
  arXiv:1704.03439 [hep-ph];
  H.~C.~Kim, M.~V.~Polyakov and M.~Praszalowicz,
  arXiv:1704.04082 [hep-ph].



  \bibitem{EARLIER}
  D.~Ebert, R.~N.~Faustov and V.~O.~Galkin,
  Phys.\ Rev.\ D {\bf 84}, 014025 (2011)
  [arXiv:1105.0583 [hep-ph]];
  D.~Ebert, R.~N.~Faustov and V.~O.~Galkin,
  Phys.\ Lett.\ B {\bf 659}, 612 (2008)
  [arXiv:0705.2957 [hep-ph]];
  W.~Roberts and M.~Pervin,
  Int.\ J.\ Mod.\ Phys.\ A {\bf 23}, 2817 (2008)
  [arXiv:0711.2492 [nucl-th]].



 \bibitem{SUMRULE}
  Z.~G.~Wang,
  Eur.\ Phys.\ J.\ C {\bf 76}, no. 2, 70 (2016)
  [arXiv:1508.01468 [hep-ph]];
  Z.~G.~Wang,
  arXiv:1704.01854 [hep-ph].



\bibitem{LATTICEC}
  M.~Padmanath and N.~Mathur,
  arXiv:1704.00259 [hep-ph].











\bibitem{ISGUR}
  E.~V.~Shuryak,
  Nucl.\ Phys.\ B {\bf 198}, 83 (1982);
  N.~Isgur and M.~B.~Wise,
  Phys.\ Rev.\ Lett.\  {\bf 66} (1991) 1130;
  A.~V.~Manohar and M.~B.~Wise,
  ``Heavy quark physics,''
  Camb.\ Monogr.\ Part.\ Phys.\ Nucl.\ Phys.\ Cosmol.\  {\bf 10}, 1 (2000).


\bibitem{MACIEK}
  M.~A.~Nowak, M.~Rho and I.~Zahed,
  Phys.\ Rev.\ D {\bf 48}, 4370 (1993)
  [hep-ph/9209272];
  M.~A.~Nowak, M.~Rho and I.~Zahed,
  Acta Phys.\ Polon.\ B {\bf 35}, 2377 (2004)
  [hep-ph/0307102].


\bibitem{BARDEEN}
  W.~A.~Bardeen and C.~T.~Hill,
  Phys.\ Rev.\ D {\bf 49} (1994) 409
  [hep-ph/9304265];
  W.~A.~Bardeen, E.~J.~Eichten and C.~T.~Hill,
  Phys.\ Rev.\ D {\bf 68}, 054024 (2003)
  [hep-ph/0305049].





\bibitem{BABAR}
  B.~Aubert {\it et al.} [BaBar Collaboration],
  Phys.\ Rev.\ Lett.\  {\bf 90}, 242001 (2003)
  [hep-ex/0304021].

\bibitem{CLEOII}
  D.~Besson {\it et al.} [CLEO Collaboration],
  Phys.\ Rev.\ D {\bf 68}, 032002 (2003)
  Erratum: [Phys.\ Rev.\ D {\bf 75}, 119908 (2007)]
  [hep-ex/0305100].


\bibitem{HOLOXX}
  J.~M.~Maldacena,
  Int.\ J.\ Theor.\ Phys.\  {\bf 38}, 1113 (1999)
  [Adv.\ Theor.\ Math.\ Phys.\  {\bf 2}, 231 (1998)]
  [hep-th/9711200];
  S.~S.~Gubser, I.~R.~Klebanov and A.~M.~Polyakov,
  Phys.\ Lett.\ B {\bf 428}, 105 (1998)
  [hep-th/9802109];
  E.~Witten,
  Adv.\ Theor.\ Math.\ Phys.\  {\bf 2}, 505 (1998)
  [hep-th/9803131];
  I.~R.~Klebanov and E.~Witten,
  Nucl.\ Phys.\ B {\bf 556}, 89 (1999)
  [hep-th/9905104].



\bibitem{HOLOXXX}
  J.~Erlich, E.~Katz, D.~T.~Son and M.~A.~Stephanov,
  Phys.\ Rev.\ Lett.\  {\bf 95}, 261602 (2005)
  [hep-ph/0501128];
  L.~Da Rold and A.~Pomarol,
  Nucl.\ Phys.\ B {\bf 721}, 79 (2005)
  [hep-ph/0501218].





\bibitem{HOLOXXXX}
  S.~Hong, S.~Yoon and M.~J.~Strassler,
  JHEP {\bf 0604}, 003 (2006)
  [hep-th/0409118];
  J.~Erlich, G.~D.~Kribs and I.~Low,
  Phys.\ Rev.\ D {\bf 73}, 096001 (2006)
  doi:10.1103/PhysRevD.73.096001
  [hep-th/0602110];
  H.~R.~Grigoryan and A.~V.~Radyushkin,
  Phys.\ Rev.\ D {\bf 76}, 095007 (2007)
  [arXiv:0706.1543 [hep-ph]];
  H.~R.~Grigoryan and A.~V.~Radyushkin,
  Phys.\ Lett.\ B {\bf 650}, 421 (2007)
  [hep-ph/0703069];
  S.~S.~Afonin and I.~V.~Pusenkov,
  EPJ Web Conf.\  {\bf 125}, 04004 (2016)
  [arXiv:1606.06091 [hep-ph]];
  N.~R.~F.~Braga, M.~A.~Martin Contreras and S.~Diles,
  Europhys.\ Lett.\  {\bf 115}, no. 3, 31002 (2016)
  [arXiv:1511.06373 [hep-th]];
  A.~Gorsky, S.~B.~Gudnason and A.~Krikun,
  Phys.\ Rev.\ D {\bf 91}, no. 12, 126008 (2015)
  [arXiv:1503.04820 [hep-th]].






\bibitem{SSX}
  T.~Sakai and S.~Sugimoto,
  Prog.\ Theor.\ Phys.\  {\bf 113}, 843 (2005)
  [hep-th/0412141];
  T.~Sakai and S.~Sugimoto,
  Prog.\ Theor.\ Phys.\  {\bf 114}, 1083 (2005)
  [hep-th/0507073].


\bibitem{HIDDEN}
  T.~Fujiwara, T.~Kugo, H.~Terao, S.~Uehara and K.~Yamawaki,
  Prog.\ Theor.\ Phys.\  {\bf 73}, 926 (1985).

  \bibitem{LIUHEAVY}
  Y.~Liu and I.~Zahed,
  Phys.\ Rev.\ D {\bf 95}, no. 5, 056022 (2017)
  [arXiv:1611.03757 [hep-ph]].
  Y.~Liu and I.~Zahed,
  arXiv:1611.04400 [hep-ph];
  Y.~Liu and I.~Zahed,
  arXiv:1704.03412 [hep-ph].




\bibitem{CHIN}
  S.~w.~Li,
  Phys.\ Rev.\ D {\bf 96}, no. 10, 106018 (2017)
  [arXiv:1707.06439 [hep-th]];
  W.~Cai and S.~w.~Li,
  Eur.\ Phys.\ J.\ C {\bf 78}, no. 6, 446 (2018)
  [arXiv:1712.06304 [hep-th]];
  S.~w.~Li,
  Phys.\ Rev.\ D {\bf 99}, no. 4, 046013 (2019)
  [arXiv:1812.03482 [hep-th]].




\bibitem{SSXB}
  H.~Hata, T.~Sakai, S.~Sugimoto and S.~Yamato,
  Prog.\ Theor.\ Phys.\  {\bf 117} (2007) 1157
  [hep-th/0701280 [HEP-TH]].


\bibitem{SSXBB}
  K.~Hashimoto, T.~Sakai and S.~Sugimoto,
  Prog.\ Theor.\ Phys.\  {\bf 120} (2008) 1093
  [arXiv:0806.3122 [hep-th]];
  K.~Y.~Kim and I.~Zahed,
  JHEP {\bf 0809}, 007 (2008)
  [arXiv:0807.0033 [hep-th]].










  \bibitem{CSLIGHT}
  H.~Hata and M.~Murata,
  Prog.\ Theor.\ Phys.\  {\bf 119}, 461 (2008)
  [arXiv:0710.2579 [hep-th]].



\bibitem{KOJI}
  K.~Hashimoto, N.~Iizuka, T.~Ishii and D.~Kadoh,
 Phys.\ Lett.\ B {\bf 691}, 65 (2010)
[arXiv:0910.1179 [hep-th]].


  \bibitem{CSTHREE}
  P.~H.~C.~Lau and S.~Sugimoto,
  arXiv:1612.09503 [hep-th].






  \bibitem{SKYRME}
  I.~Zahed and G.~E.~Brown,
  Phys.\ Rept.\  {\bf 142}, 1 (1986);
   Multifaceted Skyrmion, Eds. M.~Rho and I.~Zahed, World Scientific, 2016.






\bibitem{SKYRMEHEAVY}
  N.~N.~Scoccola,
  Nucl.\ Phys.\ A {\bf 532}, 409C (1991);
  M.~Rho, D.~O.~Riska and N.~N.~Scoccola,
  Z.\ Phys.\ A {\bf 341}, 343 (1992);
  D.~P.~Min, Y.~s.~Oh, B.~Y.~Park and M.~Rho,
  hep-ph/9209275.
  Y.~s.~Oh, B.~Y.~Park and D.~P.~Min,
  Phys.\ Rev.\ D {\bf 49}, 4649 (1994)
  [hep-ph/9402205];
  Y.~s.~Oh, B.~Y.~Park and D.~P.~Min,
  Phys.\ Rev.\ D {\bf 50}, 3350 (1994)
  [hep-ph/9407214];
  D.~P.~Min, Y.~s.~Oh, B.~Y.~Park and M.~Rho,
  Int.\ J.\ Mod.\ Phys.\ E {\bf 4}, 47 (1995)
  [hep-ph/9412302];
  Y.~s.~Oh and B.~Y.~Park,
  Phys.\ Rev.\ D {\bf 51}, 5016 (1995)
  [hep-ph/9501356];
  J.~Schechter, A.~Subbaraman, S.~Vaidya and H.~Weigel,
  Nucl.\ Phys.\ A {\bf 590}, 655 (1995)
  Erratum: [Nucl.\ Phys.\ A {\bf 598}, 583 (1996)]
  [hep-ph/9503307];
  Y.~s.~Oh and B.~Y.~Park,
  Z.\ Phys.\ A {\bf 359}, 83 (1997)
  [hep-ph/9703219];
  C.~L.~Schat and N.~N.~Scoccola,
  Phys.\ Rev.\ D {\bf 61}, 034008 (2000)
  [hep-ph/9907271];
  N.~N.~Scoccola,
  arXiv:0905.2722 [hep-ph];
  N.~Itzhaki, I.~R.~Klebanov, P.~Ouyang and L.~Rastelli,
  Nucl.\ Phys.\ B {\bf 684}, 264 (2004)
  [hep-ph/0309305];
  J.~P.~Blanckenberg and H.~Weigel,
  Phys.\ Lett.\ B {\bf 750}, 230 (2015)
  [arXiv:1505.06655 [hep-ph]];
  M.~Praszalowicz,
  PoS CORFU {\bf 2017}, 025 (2018)
  [arXiv:1805.07729 [hep-ph]].


  \bibitem{FEWX}
  A.~Paredes and P.~Talavera,
  Nucl.\ Phys.\ B {\bf 713}, 438 (2005)
  [hep-th/0412260];
  J.~Erdmenger, N.~Evans and J.~Grosse,
  JHEP {\bf 0701}, 098 (2007);
  [hep-th/0605241].
  J.~Erdmenger, K.~Ghoroku and I.~Kirsch,
  JHEP {\bf 0709} (2007) 111
  [arXiv:0706.3978 [hep-th]];
  C.~P.~Herzog, S.~A.~Stricker and A.~Vuorinen,
  JHEP {\bf 0805}, 070 (2008)
  [arXiv:0802.2956 [hep-th]];
  Y.~Bai and H.~C.~Cheng,
  JHEP {\bf 1308}, 074 (2013)
  [arXiv:1306.2944 [hep-ph]];
  K.~Hashimoto, N.~Ogawa and Y.~Yamaguchi,
  JHEP {\bf 1506}, 040 (2015)
  [arXiv:1412.5590 [hep-th]].





\bibitem{BRODSKY}
  G.~F.~de Teramond, S.~J.~Brodsky, A.~Deur, H.~G.~Dosch and R.~S.~Sufian,
  arXiv:1611.03763 [hep-ph];
  H.~G.~Dosch, G.~F.~de Teramond and S.~J.~Brodsky,
  Phys.\ Rev.\ D {\bf 92} (2015) no.7,  074010
  [arXiv:1504.05112 [hep-ph]];
H.~G.~Dosch, G.~F.~de Teramond and S. J. Brodsky,
  Phys.\ Rev. \ D {\bf 95} (2017) no. 3, 034016
  [arXiv:1612.02370 [hep-ph]].




\bibitem{COBIX}
  J.~Sonnenschein and D.~Weissman,
  Nucl.\ Phys.\ B {\bf 920} (2017) 319
  [arXiv:1606.02732 [hep-ph]];
  J.~Sonnenschein and D.~Weissman,
  arXiv:1812.01619 [hep-ph].






\bibitem{LS}
  M.~Luscher,
  Phys.\ Lett.\  {\bf 70B}, 321 (1977).
  B.~M.~Schechter,
  Phys.\ Rev.\ D {\bf 16}, 3015 (1977).



\bibitem{EXPLO1}
  D.~M.~Ostrovsky, G.~W.~Carter and E.~V.~Shuryak,
  Phys.\ Rev.\ D {\bf 66}, 036004 (2002)
  [hep-ph/0204224].


\bibitem{EXPLO2}
  E.~Shuryak and I.~Zahed,
  Phys.\ Rev.\ D {\bf 67}, 014006 (2003)
  [hep-ph/0206022].
















\bibitem{ANTON}
  F.~Brunner, D.~Parganlija and A.~Rebhan,
  Phys.\ Rev.\ D {\bf 91}, no. 10, 106002 (2015)
  Erratum: [Phys.\ Rev.\ D {\bf 93}, no. 10, 109903 (2016)]
  [arXiv:1501.07906 [hep-ph]].







\bibitem{RG}
  K.~M.~Case,
  Phys.\ Rev.\  {\bf 80}, 797 (1950);
  S.~R.~Beane, P.~F.~Bedaque, L.~Childress, A.~Kryjevski, J.~McGuire and U.~van Kolck,
  Phys.\ Rev.\ A {\bf 64}, 042103 (2001)
  [quant-ph/0010073];
  E.~Braaten and D.~Phillips,
  Phys.\ Rev.\ A {\bf 70}, 052111 (2004)
  [hep-th/0403168].






\bibitem{EFIMOV}
  V.~Efimov,
  Phys.\ Lett.\  {\bf 33B}, 563 (1970).




\bibitem{MALT}
  A.~Francis, R.~J.~Hudspith, R.~Lewis and K.~Maltman,
  arXiv:1810.10550 [hep-lat].



\bibitem{SW}
  M.~J.~Savage and M.~B.~Wise,
  Phys.\ Lett.\ B {\bf 248}, 177 (1990).

\bibitem{CALLANKLEB}
C. G. Callan and I. Klebanov,  Nucl. Phys. {\bf B262} (1985) 365.  For a review, see e.g. M.A. Nowak, M. Rho and I. Zahed, {\it Nuclear Chiral Dynamics}, World Scientific, 
Singapore (1996). 

\bibitem{NEW}
  Y.~R.~Liu, H.~X.~Chen, W.~Chen, X.~Liu and S.~L.~Zhu,
  arXiv:1903.11976 [hep-ph].


\bibitem{KIRITSIS}
  U.~Gursoy and E.~Kiritsis,
  JHEP {\bf 0802}, 032 (2008)
  [arXiv:0707.1324 [hep-th]];
  U.~Gursoy, E.~Kiritsis and F.~Nitti,
  JHEP {\bf 0802}, 019 (2008)
  [arXiv:0707.1349 [hep-th]].













\end{thebibliography}
\end{document}